\documentclass[12pt, a4paper]{article}
\usepackage{graphicx}
\usepackage{subeqnarray}
\begin{document}
%
%                         _________________
%                         |               |
%                         |               |
%                         |  VERSIONE  C  |
%                         |               |
%                         |  0000.0000    |
%                         |               |
%                         |  xxxxx        |
%                         _________________
%
%
%
\def\astrobj#1{#1}
\newenvironment{lefteqnarray}{\arraycolsep=0pt\begin{eqnarray}}
{\end{eqnarray}\protect\aftergroup\ignorespaces}
\newenvironment{lefteqnarray*}{\arraycolsep=0pt\begin{eqnarray*}}
{\end{eqnarray*}\protect\aftergroup\ignorespaces}
\newenvironment{leftsubeqnarray}{\arraycolsep=0pt\begin{subeqnarray}}
{\end{subeqnarray}\protect\aftergroup\ignorespaces}
\newcommand{\diff}{{\rm\,d}}
\newcommand{\img}{{\rm i}}
\newcommand{\sV}{\mskip 3mu /\mskip-10mu V}
\newcommand{\sP}{\mskip 3mu /\mskip-10mu p}
\newcommand{\sT}{\mskip 3mu /\mskip-08mu T}
\newcommand{\sX}{\mskip 3mu /\mskip-12mu X}
\newcommand{\appleq}{\stackrel{<}{\sim}}
\newcommand{\appgeq}{\stackrel{>}{\sim}}
\newcommand{\Int}{\mathop{\rm Int}\nolimits}
\newcommand{\Nint}{\mathop{\rm Nint}\nolimits}
\newcommand{\arcsinh}{\mathop{\rm arcsinh}\nolimits}
\newcommand{\range}{{\rm -}}
\newcommand{\sgn}{\mathop{\rm sgn}\nolimits}
\newcommand{\displayfrac}[2]{\frac{\displaystyle #1}{\displaystyle #2}}
\def\astrobj#1{#1}
%\begin{titlepage}
%\setcounter{page}{0}
%\headnote{Astron.~Nachr.~000 (2001) 0, 000--000}
%\makeheadline
%
\title{Elliptical galaxies: rotationally distorted, after all}
\author{{R.~Caimmi}\footnote{
{\it Astronomy Department, Padua Univ., Vicolo Osservatorio 2,
I-35122 Padova, Italy}
email: roberto.caimmi@unipd.it~~~
fax: 39-049-8278212}
%
%, {T.~Valentinuzzi}\footnote{
%{\it Astronomy Department, Padua Univ., Vicolo Osservatorio 2,
%I-35122 Padova, Italy}
%email: tiziano.valentinuzzi@unipd.it~~~
%fax: 39-049-8278212}
%
\phantom{agga}}
%
%\medskip
%\small{Dipartimento di Astronomia}}
%
%\date{Received..................................................
%Accepted..................................................}
\maketitle
\begin{quotation}
\section*{}
\begin{Large}
\begin{center}
%\summary

Abstract

\end{center}
\end{Large}
\begin{small}

\noindent\noindent

On the basis of earlier investigations on
homeoidally striated Mac Laurin spheroids
and Jacobi ellipsoids (Caimmi and Marmo
2005; Caimmi 2006a, 2007), different sequences
of configurations are defined and represented
on the ellipticity-rotation plane,
$({\sf O}\hat{e}\chi_v^2)$.   The rotation
parameter, $\chi_v^2$, is defined as the ratio,
$E_{\rm rot}/E_{\rm res}$, of kinetic energy
related to the mean tangential equatorial velocity
component, $M(\overline{v_\phi})^2/2$, to kinetic
energy related to tangential equatorial component
velocity dispersion, $M\sigma_{\phi\phi}^2/2$, and
residual motions, $M(\sigma_{ww}^2+\sigma_{33}^2)/2$.
Without loss of generality (above a threshold
in ellipticity values), the analysis is restricted to
systems with isotropic stress tensor, which
may be considered as adjoint configurations
to any assigned homeoidally striated density
profile with anisotropic stress tensor, different
angular momentum, and equal remaining parameters.
The description of configurations on the
$({\sf O}\hat{e}\chi_v^2)$ plane is extended
on two respects, namely (a) from equilibrium
to nonequilibrium figures, where the virial
equations hold with additional kinetic energy,
and (b) from real to imaginary rotation, where
the effect is elongating instead of flattening,
with respect to the rotation axis.
%
%The key concept is that distorted boundaries are due
%to additional kinetic energy of tangential
%equatorial velocity component, with respect
%to nonrotating spherical configurations,
%regardless of the angular momentum and the
%stress tensor.
%
An application is made to
a subsample $(N=16)$ of elliptical galaxies
extracted from richer samples $(N=25,~N=48)$
of early type galaxies investigated within the
SAURON project (Cappellari et al. 2006, 2007).
Sample objects are idealized as homeoidally
striated MacLaurin
spheroids and Jacobi ellipsoids, and their
position on the $({\sf O}\hat{e}\chi_v^2)$
plane is inferred from observations following
a procedure outlined in an earlier paper
(Caimmi 2009).   The position of related
adjoint configurations with isotropic stress
tensor is also determined.   With a single
exception (NGC 3379), slow rotators are
characterized by low ellipticities $(0\le
\hat{e}<0.2)$, low anisotropy parameters
$(0\le\delta<0.15)$, and low rotation
parameters $(0\le\chi_v^2<0.15)$, while fast
rotators show large ellipticities $(0.2\le
\hat{e}<0.65)$, large anisotropy parameters
$(0.15\le\delta<0.35)$, and large rotation
parameters $(0.15\le\chi_v^2<0.5)$.   An
alternative kinematic classification with
respect to earlier attempts (Emsellem et
al. 2007) needs richer samples for providing
additional support to the above mentioned
results.   A possible interpretation of slow
rotators as nonrotating at all and elongated
due to negative anisotropy parameters,
instead of flattened due to positive
anisotropy parameters, is exploited.
Finally, the elliptical side of the Hubble
morphological sequence is interpreted as a
sequence of equilibrium (adjoint) configurations
where the ellipticity is an increasing function
of the rotation parameter, slow rotators
correspond to early classes (E0-E2 in the
oblate limit and E$-$2-E0 in the prolate limit)
and fast rotators to late classes (E3-E6).
In this view, boundaries are rotationally
distorted regardless of angular momentum
and stress tensor, where rotation has to
be intended as due to additional kinetic energy
of tangential equatorial velocity components,
with respect to spherical configurations
with isotropic stress tensor.

\noindent
{\it keywords -
galaxies: kinematics and dynamics - galaxies: elliptical.}
%END
%\end{titlepage}
\end{small}
\end{quotation}

\section{Introduction} \label{intro}

According to their original classification (Hubble 1926),
elliptical galaxies are subdivided into eight classes,
designated E0, E1, ..., E7, the numerical index being
the integer nearest to $10\hat{e}$, where $\hat{e}$ is
the ellipticity of related galaxies as projected on the
sky, and $0\le\hat{e}\le0.7$ is inferred from observations%
\footnote{
In modern classifications, galaxies within the class E7
are considered as lenticulars instead of ellipticals (for
further details refer to e.g., Caimmi 2006b), but it is
unrelevant to the aim of the current investigation.   For
this reason, Hubble's original classification shall be
considered in the following.}.
Hubble's original scheme has generally been considered
satisfactory, but the physical explanation of the distorted
boundary has been changed in time.

The classical interpretation starts from the evidence that
(i) all celestial bodies are known to be in rotation, and
(ii) the symmetry of figure shown by elliptical galaxies is
precisely such as rotation might be expected to produce;
which, in turn, suggests (iii) an inquiry as to how far the
observed figures of elliptical galaxies can be explained as
the figures assumed by masses rotating under their own
gravitation.   In this view, boundaries are distorted by
systematic rotation around the minor axis i.e. angular
momentum parallel to the rotation axis.   For further
details refer to the parent textbook (Jeans 1929, Chap.\,XIII,
\S\S299-302) and to recent attempts (e.g., Caimmi 2006b).

About fourty years ago, observations begun to yield increasing
evidence that (giant) elliptical galaxies cannot be sustained
by systematic rotation (e.g., Bertola and Capaccioli 1975;
Binney 1976; Illingworth 1977, 1981; Schechter and Gunn 1979).
Accordingly, (giant) elliptical galaxies were conceived as
systems with triplanar symmetry and ellipsoidal boundaries
set up by specific anisotropic stress tensor (Binney 1976,
1978, 1980) and a negligible contribution from angular
momentum.   Owing to high-resolution simulations, the same
holds also for (nonbaryonic) dark matter haloes hosting
galaxies and clusters of galaxies (e.g., Hoeft et al. 2004;
Rasia et al. 2004; Bailin and Steinmetz 2004).   In this
view, boundaries are distorted by systematic rotation
around the minor axis and/or anisotropic stress tensor.

It is worth mentioning that an anisotropic stress tensor
does not imply orbital anisotropy and vice versa.   With
regard to e.g., nonrotating systems, and leaving aside
instabilities, orbits could be only radial or only
circular, and the shape maintain spherical or flat.
On the other hand, anisotropic stress tensors yield
distorted boundaries (e.g., Binney 1976, 2005, the
last hereafter quoted as B05).
The extent to which angular momentum and anisotropic
stress tensor are effective in determining the shape of
a system, may be quantified by the ratio, $\chi_v^2$, of
kinetic energy due to figure rotation to kinetic energy
due to remaining motions.   It is, in turn, related to a
squared velocity ratio, $(\chi_v^2)_{\rm obs}$, inferred
from observations.   For assigned density profiles, the
dependence of the rotation parameter on the shape may
be determined using the tensor virial theorem.

In the special case of homeoidally striated ellipsoids,
where the isopycnic (i.e. constant density) surfaces
are similar and similarly placed (Roberts 1962), the
rotation parameter, $\chi_v$, has been explicitly
expressed as a function of the ellipticity, $\hat{e}$,
restricting to oblate shapes (Binney 1976).   Configurations
with assigned anisotropic stress tensor are represented
on the $({\sf O}\hat{e}\chi_v)$ plane as a family of
curves which branch off from the origin, $(\hat{e},\chi_v)
=(0,0)$, and increase for increasing ellipticity.   The
curve related
to the isotropic stress tensor makes an upper limit, while
their counterparts characterized by an anisotropic stress
tensor $(\sigma_{11}=\sigma_{22}>\sigma_{33})$ lie below
(Binney 1976).

The tidal action from an embedding, nonbaryonic dark halo
makes the shape of the embedded elliptical galaxy, closer
to its own.   In other words, a less flattened dark halo
makes a less flattened embedded elliptical galaxy, and
vice versa.   Accordingly, a family of curves depending on two
parameters, namely the fractional mass, $m=M_{\rm halo}/
M_{\rm egal}$, and the axis ratio of the dark halo
(supposed to be axisymmetric), $\epsilon_{\rm halo}$,
can fit to the data on the $({\sf O}\hat{e}\chi_v)$ plane
even if the stress tensor is isotropic (Caimmi 1992).

The rotation parameter, $\chi_v$, is independent of the
density profile for homeoidally striated ellipsoids
(Roberts 1962; Binney 1976) which, in general, make
only a first approximation to equilibrium configurations
(e.g., Vandervoort 1980b; Vandervoort and Welty 1981;
Lai et al. 1993).    With regard to equilibrium
configurations e.g., polytropes (Vandervoort 1980a),
the rotation parameter, $\chi_v$, depends on the
density profile (Caimmi 1980).   Accordingly, a
family of curves depending on a single parameter,
namely the polytropic index, $n$, can fit to the
data on the $({\sf O}\hat{e}\chi_v)$ plane even if
the stress tensor is isotropic (Vandervoort 1980a;
Caimmi 1983).

Sysytematic rotation and anisotropic stress tensor
are effective to the same extent in flattening or
elongating a spherical shape, provided imaginary
rotation (around the major axis) is considered in
the latter case (Caimmi 1996b, 2007, 2008; the last
quoted hereafter as C07, C08, respectively).   The
key concept is that the distribution function is
independent on the sign of tangential velocity
components, and the whole set of possible
configurations is characterized by an equal
amount of both kinetic and potential energy
(Lynden-Bell 1960, 1962; Meza 2002).

In other
words, clockwise and counterclockwise tangential
velocity components are indistinguishable to
this respect, and no change in shape occurs for
any configuration between the limiting cases:
(i) equal amount of clockwise and counterclockwise
tangential equatorial velocity components, which
implies a null mean value; (ii) only clockwise
or counterclockwise tangential equatorial velocity
components, which implies a maximum or minimum
(according to the sign) mean value.   In particular,
a single configuration exists with isotropic stress
tensor, to be conceived as ``adjoint'' to its
generic counterpart with anisotropic stress tensor.
The above considerations hold above a threshold in
ellipticity $(\hat{e}\ge\hat{e}_{\rm adj})$ which,
on the other hand, lies below the threshold for
elliptical galaxies $(\hat{e}\ge\hat{e}_{\rm ell}
\ge\hat{e}_{\rm adj})$.   Accordingly, elliptical
galaxies always admit adjoints configurations
with isotropic stress tensor.

The tensor virial theorem provides a rigorous global
link between angular momentum, anisotropic stress
tensor and shape, but the quantities appearing therein
cannot easily be deduced from observations.   A
classical approximation to the rotation parameter,
$\chi_v$, is $(\chi_v)_{\rm obs}=V_{\rm max}/\sigma_0$,
where $V_{\rm max}$ is the peack rotation velocity and
$\sigma_0$ a centrally averaged velocity dispersion,
both projected on the line of sight (e.g., Illingworth
1977, 1981; Schechter and Gunn 1979).   With the advent
of integral-field spectroscopy (Bacon et al. 2002; de
Zeeuw et al. 2002; Kelz et al. 2003) a rigorous
connection can be established between the tensor virial
theorem and observations, provided the former is
reformulated in terms of sky-averages.   In particular,
the rotation parameter inferred from the data is
$(\chi_v)_{\rm obs}=<\widetilde{v}_\|^2>^{1/2}/
<\widetilde{\sigma}_\|^2>^{1/2}$, where $\widetilde
{v}_\|$ is the mean velocity projected on the line of
sight, $\widetilde{\sigma}_\|^2$ is the related variance,
and $<\widetilde{v}_\|^2>$, $<\widetilde{\sigma}_\|^2>$,
are sky-averages with respect to the mass (B05).

In general, the expression of the rotation
parameter depends on what is intended as
rotation energy.   Usually, the kinetic
energy is decomposed into contributions
from ordered and random motions (e.g.,
B05).   In the current attempt, the kinetic
energy shall be decomposed into contributions
from cylindrical rotation, $M(\overline{v_
\phi})^2/2$, and tangential equatorial component
velocity dispersion, $M\sigma_{\phi\phi}^2/2$,
plus residual motions, $M(\sigma_{ww}^2+
\sigma_{33}^2)/2$; the rotation parameter,
$\chi_v^2$, shall be defined as the ratio of the
first to the sum of the remaining two above
mentioned terms.

In this view, an extension to nonequilibrium
configurations and imaginary rotation appears
quite natural, in the light of a general theory
where boudaries are distorted by angular momentum
and/or anisotropic stress tensor (C07, C08)
and triaxial configurations due to the
occurrence of bifurcation points are also considered
(Caimmi 1996a,b, 2006a,b; CM05).   In
dealing with imaginary tangential equatorial
velocity components, the rotation
parameter, $\chi_v^2$, has to be used instead of 
$\chi_v$, where positive and negative values are
related to real and imaginary angular momentum,
respectively.

The present investigation is mainly devoted to the
following points: (i) representation of nonequilibrium
figures on the $({\sf O}\hat{e}\chi_v^2)$ plane,
restricted to homeoidally striated MacLaurin spheroids
and Jacobi ellipsoids and edge-on orientations with
the major axis perpendicular to the line of sight;
(ii) location on the $({\sf O}\hat{e}\chi_v^2)$ plane
of elliptical galaxies with assigned inclination angle
and anisotropy parameter, restricted to axisymmetric
shapes and classified as fast and slow rotators
(Cappellari et al. 2007; hereafter quoted as S\,X);
(iii) extent to which nonequilibrium figures fit to
elliptical galaxies; (iv) interpretation of slow rotators
as intrinsically prolate, nonrotating bodies, and
related location on the $({\sf O}\hat{e}\chi_v^2)$ plane;
(v) representation of adjoint configurations in imaginary
rotation with isotropic stress tensor.

The work is organized as follows.   The rotation
parameter, $\chi_v^2$, is expressed as a function
of the intrinsic meridional ellipticity, $\hat{e}$,
for equilibrium and nonequilibrium figures with
isotropic stress tensor, in Section \ref{sonf}.   The position
of elliptical galaxies on the $({\sf O}\hat{e}\chi_v^2)$
plane, with regard to a restricted sample for which
fiducial values of the inclination angle and the
anisotropy parameter can be assigned, together with
their adjoint configurations where the stress tensor
is isotropic, is determined in Section \ref{cowo}.
A comparison with earlier attempts is made in
Section \ref{chCB}.   The
possibility that slow rotators are in fact prolate,
nonrotating bodies, is exploited in Section \ref{disc},
where an interpretation of the elliptical side of
the Hubble sequence is also proposed.
The conclusion is drawn in Section \ref{conc}.

\section{Nonequilibrium figures}\label{sonf}

Given a mass distribution with assigned rotation axis,
$x_3$, the kinetic energy is usually decomposed into
contributions from ordered and random motion (e.g.,
B05), where the former arises from mean velocity
components within any infinitesimal volume element,
and the latter from related variances which appear
in the expression of the stress tensor.   In this
view, the contribution of streaming motion to velocity
dispersion (along a selected direction) is not included
in the stress tensor.

A different approach shall be
exploited in the current attempt.   More specifically,
the kinetic energy shall be conceived as the sum
of two contributions: one, related to either the
mass-weighted tangential equatorial velocity component, $M(\overline
{v_\phi})^2/2$, or the moment-of-inertia-weigthted
angular velocity, $I_3(\overline{\Omega})^2/2$, and one
other to the contribution of the related
velocity dispersion plus the remaining velocity components,
radial equatorial, $\overline{(v_w^2)}$, and radial
polar, $\overline{(v_3^2)}$.   The above contributions
shall be hereafter quoted as rotation kinetic energy,
$E_{\rm rot}$, and residual kinetic energy, $E_{\rm res}$,
respectively.   In this view, the contribution of
streaming motion to velocity dispersion (along a
selected direction) is included in the stress tensor.

It is worth mentioning that the mass-weighted tangential
equatorial velocity component and the moment-of-inertia-weigthted
angular velocity yield different kinetic energy values
(for a formal discussion refer to Appendix \ref{a:ciro}).
As the current investigation deals with linear velocities,
only the former shall be considered here.

If the tensor virial theorem holds for only time-averaged
quantities, the related mass distribution can no longer be
considered in dynamical or hydrostatic equilibrium, but still
in virial equilibrium (C07), which makes a special kind of
nonequilibrium figures.   In the following, ``nonequilibrium''
has to be intended as ``far from dynamical or hydrostatic
equilibrium but still in virial equilibrium''.   For
nonequilibrium figures, the tensor virial equations must
be related to an adjoint configuration with equal density
profile and rotation energy, but different residual-energy
tensor components, $(\widetilde{E}_{\rm res})_{pp}=\zeta_
{pp}E_{\rm res}$.   The result is (C07):
\begin{leftsubeqnarray}
\slabel{eq:virga}
&& 2(E_{\rm rot})_{qq}+2\zeta_{qq}E_{\rm res}+(E_{\rm pot})_{qq}=0~~;\qquad
q=1,2~~; \\
\slabel{eq:virgb}
&& \phantom{q2(E_{\rm rot})_{qq}+}2\zeta_{33}E_{\rm res}+(E_{\rm pot})_{33}=
0~~;
\label{seq:virg}
\end{leftsubeqnarray}
where $(E_{\rm pot})_{pq}$ is the potential-energy tensor and the
coefficients, $\zeta_{pp}$, may be understood as
generalized anisotropy parameters (CM05, Caimmi
2006a,b; C07).
The combination of Eqs.\,(\ref{eq:virga}) and
(\ref{eq:virgb}) yields:
\begin{leftsubeqnarray}
\slabel{eq:virpa}
&& 2(E_{\rm rot})_{qq}-\frac{\zeta_{qq}}{\zeta_{33}}(E_{\rm pot})_{33}+
(E_{\rm pot})_{qq}=0~~;\qquad q=1,2~~; \\
\slabel{eq:virpb}
&& 2E_{\rm res}=-\frac1{\zeta_{33}}(E_{\rm pot})_{33}~~;
\label{seq:virp}
\end{leftsubeqnarray}
and the potential-energy tensor for homeoidally
striated Jacobi ellipsoids reads (CM05):
\begin{leftsubeqnarray}
\slabel{eq:Epota}
&& (E_{\rm pot})_{pq}=-\frac{GM^2}{a_1}\nu_{\rm sel}(B_{\rm sel})_{pq}~~;
\qquad p,q=1,2,3~~; \\
\slabel{eq:Epotb}
&& (B_{\rm sel})_{pq}=\delta_{pq}\epsilon_{p2}\epsilon_{p3}A_p~~;\qquad
B_{\rm sel}=\sum_{s=1}^3\epsilon_{s2}\epsilon_{s3}A_s~~;
\label{seq:Epot}
\end{leftsubeqnarray}
where $\delta_{pq}$ is the Kronecker symbol, $G$ the
constant of gravitation, $\nu_{\rm sel}$ a profile
factor (i.e. depending only on the density profile),
$a_p$ are semiaxes, $\epsilon_{pq}=a_p/a_q$ axis ratios,
$A_p$ shape factors (i.e. depending only on the axis
ratios).

The substitution of Eqs.\,(\ref{seq:Epot}), (\ref{seq:Ekpp}),
(\ref{seq:Ekin}), into (\ref{seq:virp}) yields after some
algebra:
\begin{leftsubeqnarray}
\slabel{eq:vespa}
&& (\overline{v_\phi})^2=\frac{GM}{a_1}\nu_{\rm sel}\left[B_{\rm sel}-
\frac{\zeta}{\zeta_{33}}(B_{\rm sel})_{33}\right]~~; \\
\slabel{eq:vespb}
&& \sigma_{\phi\phi}^2+\sigma_{ww}^2+\sigma_{33}^2=\frac{GM}{a_1}\nu_
{\rm sel}\frac1{\zeta_{33}}(B_{\rm sel})_{33}~~; \\
\slabel{eq:vespc}
&& \zeta=\zeta_{11}+\zeta_{22}+\zeta_{33}~~; \\
\slabel{eq:vespd}
&& \frac{\zeta}{\zeta_{pp}}=\frac{\widetilde{E}_{\rm res}/E_{\rm res}}
{(\widetilde{E}_{\rm res})_{pp}/E_{\rm res}}=\frac1{\widetilde{\zeta}_{pp}}
~~; \\
\slabel{eq:vespe}
&& \widetilde{\zeta}_{pp}=\frac{(\widetilde{E}_{\rm res})_{pp}}
{\widetilde{E}_{\rm res}}~~;
\label{seq:vesp}
\end{leftsubeqnarray}
where the anisotropy parameters, $\widetilde{\zeta}_{pp}$,
are related to equilibrium figures (CM05; C07; C08).

The rotation parameter, $\chi_v^2=E_{\rm rot}/E_{\rm res}$,
by use of Eqs.\,(\ref{seq:vesp}) and (\ref{seq:Ekin})
takes the explicit expression:
\begin{leftsubeqnarray}
\slabel{eq:chi2a}
&& \chi_v^2=\frac{(\overline{v_\phi})^2}{\sigma^2}=\zeta
\left[\widetilde{\zeta}_{33}\frac{B_{\rm sel}}
{(B_{\rm sel})_{33}}-1\right]~~; \\
\slabel{eq:chi2b}
&& \sigma^2=\sigma_{\phi\phi}^2+\sigma_{ww}^2+\sigma_{33}^2=
\sigma_{11}^2+\sigma_{22}^2+\sigma_{33}^2~~;
\label{seq:chi2}
\end{leftsubeqnarray}
which, for assigned values of the anisotropy parameters,
depends only on the axis ratios.    In an earlier attempt
(CM05) a different definition of rotation parameter has
been used, in terms of a mean squared velocity,
$<v_{\rm rot}^2>$, instead of related squared mean velocity,
$<v_{\rm rot}>^2$, which implies different anisotropy
parameters for equal shapes and rotation parameters,
in the two formulations.
The current definition is more strictly connected with
observations i.e. mean velocities and velocity dispersions.
The factor within brackets in Eq.\,(\ref{eq:chi2a})
corresponds to equilibrium figures
$(\zeta=1, \zeta_{pp}=\widetilde{\zeta}_{pp})$ while the
virial index, $\zeta=\widetilde{E}_{\rm res}/E_{\rm res}$, is
a measure of the departure from equilibrium $(0\le\zeta<
+\infty)$.

For homeoidally striated Jacobi ellipsoids, the anisotropy
parameters, $\zeta_{11}$, $\zeta_{22}$, are related
to the diagonal components of the moment-of-inertia tensor,
$I_{11}$, $I_{22}$, and a necessary and sufficient
condition for the stress tensor to be isotropic, is
$\zeta_{33}=1/3$.   In addition, bifurcation points
are independent of
the angular momentum and the stress tensor, and may be
related to the adjoint configuration where the squared
angular momentum attains a maximum or minimum value
(according to the sign) and the stress tensor
is isotropic.   This is why the effect of an anisotropic
stress tensor is equivalent to an additional real or
imaginary rotation, inducing flattening (on the equatorial
plane) or elongation (on the rotation axis), respectively.
For further details refer to the parent paper (C07) and
Appendix \ref{a:rost}.   More specifically, real or
imaginary rotation take place when the factor within
brackets in Eq.\,(\ref{eq:chi2a}) is positive or negative,
respectively.

\begin{figure*}[t]
\begin{center}
\includegraphics[scale=0.8]{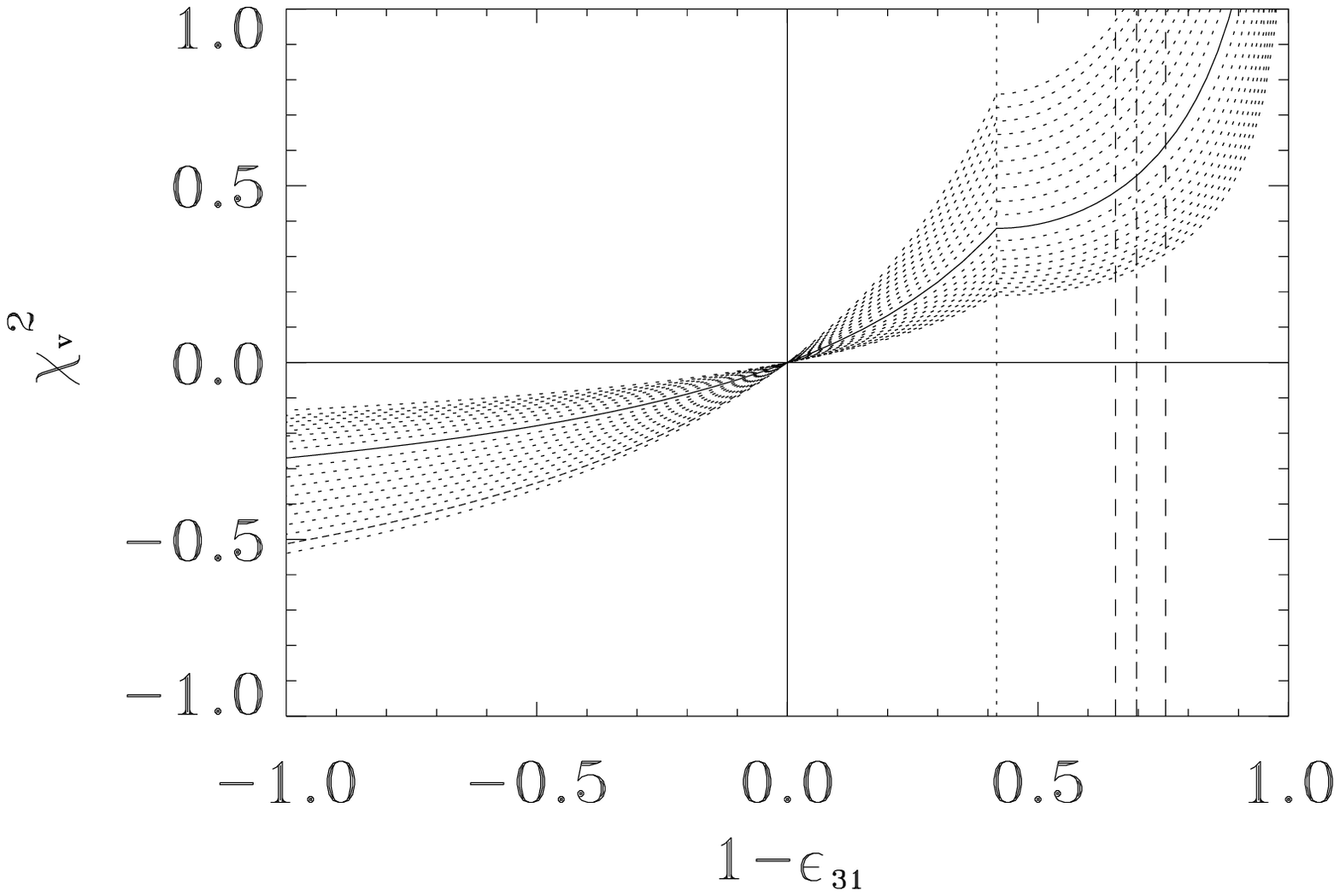}
\caption{The rotation paramerer, $\chi_v^2$, as a function
of the meridional ellipticity, $\hat{e}=1-\epsilon_{31}$,
for sequences of equilibrium (full) and nonequilibrium
(dotted) figures with isotropic stress tensor.   For
equilibrium figures, $\zeta=\widetilde{E}_{\rm res}/E_
{\rm res}=1$.   For nonequilibrium figures, $\zeta=1.1,
1.2, ..., 2.0$ (above the equilibrium sequence on the
first quadrant), and
$\zeta=1/1.1, 1/1.2, ..., 1/2.0$ (below the equilibrium 
sequence on the first quadrant).   Configurations lying on
the first and third
quadrant are in real and imaginary rotation, respectively.
Nonrotating configurations are placed at the origin.
Caption of vertical lines: dotted - bifurcation from
axisymmetric to triaxial configurations; dashed - bifurcation
from triaxial (left) and axisymmetric (right) to
pear-shaped configurations; dot-dashed - onset of
dynamical instability in axisymmetric configurations.}
\label{f:seq}
\end{center}
\end{figure*}

Given a nonequilibrium figure with assigned anisotropy
parameters, $\zeta_{pp}$, the particularization of
Eq.\,(\ref{eq:chi2a}) to the adjoint configuration with
isotropic stress tensor $(\widetilde{\zeta}_{33}=1/3)$
yields:
\begin{equation}
\label{eq:cii2}
(\chi_v^2)_{\rm iso}=\zeta\left[\frac13\frac{B_{\rm sel}}
{(B_{\rm sel})_{33}}-1\right]~~;
\end{equation}
which can be plotted as a function of the meridional
ellipticity, $\hat{e}=1-\epsilon_{31}$, where $\hat{e}
>0$ implies real rotation, $\hat{e}=0$ no rotation,
and $\hat{e}<0$ imaginary rotation.   Cases within
the range, $1/2\le\zeta\le2$, are plotted in Fig.\,\ref
{f:seq}, where the full curve corresponds to equilibrium
figures $(\zeta=1)$ and the upper on the first quadrant
$(\zeta=1.1, 1.2, ..., 2.0)$ and the lower on the first
quadrant $(\zeta=1/1.1, 1/1.2, ..., 1/2.0)$ dotted
curves to nonequilibrium figures.   Points lying on the
first and the third quadrant represent systems in real
and imaginary rotation, respectively.
The origin is the locus of nonrotating systems.   The
loci of different bifurcation points are represented
by different vertical lines, namely: from axisymmetric
to triaxial configurations (dotted); from triaxial
(left) and axisymmetric (right) to pear-shaped
configurations (dashed); and, in addition: onset of
dynamical instability in axisymmetric configurations
(dot-dashed).   The first
and the third quadrant can completely be filled with
curves expressed by Eq.\,(\ref{eq:cii2}), where the
limiting case, $\zeta=0$, corresponds to flat
configurations with no centrifugal support, and the
limiting case, $\zeta\to+\infty$, corresponds to
nonflat configurations with no pressure support.

In the general case of anisotropic stress tensor,
conformly to Eq.\,(\ref{eq:chi2a}), the trend on
the $({\sf O}\hat{e}\chi_v^2)$ plane is similar
but the curves are shifted on the right
(nonrotating flattened configurations) or on the
left (nonrotating elongated configurations),
according if $\sigma_{11}^2=\sigma_{22}^2>
\sigma_{33}^2$ or $\sigma_{11}^2=\sigma_{22}^2<
\sigma_{33}^2$, respectively.   In any case, the
adjoint equilibrium configurations with isotropic
stress tensor lie on the related sequence (full
curve in Fig.\,\ref{f:seq}) with unchanged axes,
which implies a vertical shift of a selected point
on the $({\sf O}\hat{e}\chi_v^2)$ plane, until the
adjoint configuration is attained.

\section{Comparison with observations}
\label{cowo}

Aiming in comparing model predictions to data from
observations, the position of elliptical galaxies
on the $({\sf O}\hat{e}\chi_v^2)$ plane shall be
determined.   The sample used (Caimmi and Valentinuzzi
2008; Caimmi 2009b; hereafter quoted as CV08 and C09,
respectively) is extracted from richer samples of
early-type galaxies investigated within the SAURON
project (S\,IV, S\,X).    The following parameters
may directly be deduced or derived from observations:
the effective (half-light) radius, $R_e$; the stellar
mass within the effective radius, $M_e=M(R_e)=M/2$;
the luminosity-weighted average ellipticity, on a plane
perpendicular to the line of sight, $<\hat{e}_\bot>$,
within either
an isophote enclosing an area, $\hat{A}=\pi R_e^2$,
or the largest isophote fully contained within the
SAURON field, whichever is smaller; the
luminosity-weighted squared mean velocity component,
parallel to the line of sight, $<\widetilde{v}_\|^2>$,
within either an
ellipse of area, $\hat{A}$, ellipticity, $\hat{e}_\bot$,
and related position angle, or the largest similar
ellipse fully contained within the SAURON field,
whichever is smaller; the luminosity-weighted
squared velocity dispersion, parallel to the line
of sight, $<\sigma_\|^2>$, within either an
ellipse of area, $\hat{A}$, ellipticity, $\hat{e}_\bot$,
and related position angle, or the largest similar
ellipse fully contained within the SAURON field,
whichever is smaller.   For further details refer to
the parent papers (S\,IV; S\,X) and an earlier attempt
(B05).

Two additional parameters can be inferred by fitting
the data with dynamic models.   More specifically,
the inclination angle, $i$, is deduced from the best
fitting two-integral Jeans model (S\,IV), and the
anisotropy parameter, $\delta$, is determined from
the solution of the dynamic models, supposed to be
axisymmetric (S\,X).   It is worth remembering that
fast rotators show evidence of large anisotropy and
axial symmetry, while slow rotators appear to be
nearly isotropic and moderately triaxial (S\,X).

The anisotropy parameter (e.g., B05):
\begin{leftsubeqnarray}
\slabel{eq:deltaa}
&& \delta=1-\frac{\sigma_{33}^2}{\sigma_{11}^2}=
1-\frac{\sigma_{33}^2}{\sigma_{22}^2}~~; \\
\slabel{eq:deltab}
&& \sigma_{11}=\sigma_{22}\ge\sigma_{33}~~;
\label{seq:delta}
\end{leftsubeqnarray}
may be related to the generalized anisotropy
parameters, $\zeta_{pp}$, via Eqs.\,(\ref{eq:vespd}),
(\ref{eq:vespe}), in the special case of equilibrium
figures, as:
\begin{equation}
\label{eq:delzi}
\delta=1-\frac{\zeta_{33}}{\zeta_{11}}=1-\frac{\zeta_{33}}{\zeta_{22}}=
\frac{1-3\zeta_{33}}{1-\zeta_{33}}=\frac{3\zeta_{11}-1}{\zeta_{11}}=
\frac{3\zeta_{22}-1}{\zeta_{22}}~~;
\end{equation}
which is restricted to homeoidally striated
MacLaurin spheroids.

For representing sample objects on the $({\sf O}
\hat{e}\chi_v^2)$ plane, intrinsic values of
velocities and ellipticities with regard to
edge-on configurations have to be used, instead
of projected values along the line of sight.
The corrections for edge-on (edo) configurations
are (Binney and Tremaine 1987, Chap.\,4, \S4.3):
\begin{lefteqnarray}
\label{eq:epin}
&& 1-\epsilon_{31}^2=\frac{1-<\epsilon_\bot>^2}{\sin^2i}~~; \\
\label{eq:elin}
&& \hat{e}=1-\epsilon_{31}=1-\left[1-\frac{1-
<\hat{e}_\bot>(2-<\hat{e}_\bot>)}{\sin^2i}\right]^{1/2}~~;
\end{lefteqnarray}
for the meridional axis ratio and ellipticity,
where $<\epsilon_\bot>$ and $<\hat{e}_\bot>$
are the luminosity averaged axis ratio and
ellipticity related to an inclination angle,
$i$, between the symmetry axis and the line
of sight ($i=90^\circ$ for edge-on configurations),
and:
\begin{lefteqnarray}
\label{eq:vrin}
&& [<\widetilde{v}_\|^2>]_{\rm edo}=\frac{[<\widetilde{v}_\|^2>]_{\rm obs}}
{\sin^2i}~~; \\
\label{eq:sgin}
&& [<\sigma_\|^2>]_{\rm edo}=\frac{[<\sigma_\|^2>]_{\rm obs}}
{1-\delta\cos^2i}~~;
\end{lefteqnarray}
under the assumption of axisymmetric $(a_1=a_2)$
configurations with axisymmetric $(\sigma_{11}=
\sigma_{22})$ stress tensor (S\,X).

The intrinsic squared mean rotational velocity
and squared velocity dispersion are:
\begin{lefteqnarray}
\label{eq:vfe2}
&& (\overline{v_\phi})^2=2[<\widetilde{v}_\|^2>]_{\rm edo}~~; \\
\label{eq:sge2}
&& \sigma^2=\sigma_{11}^2+\sigma_{22}^2+\sigma_{33}^2=2[<\sigma_\|^2>]_
{\rm edo}+(1-\delta)[<\sigma_\|^2>]_{\rm edo}\nonumber \\
&& \phantom{\sigma^2=\sigma_{11}^2+\sigma_{22}^2+\sigma_{33}^2}
=(3-\delta)[<\sigma_\|^2>]_{\rm edo}~~;
\end{lefteqnarray}
in terms of edge-on mean rotational velocity and
velocity dispersion, and:
\begin{lefteqnarray}
\label{eq:vfo2}
&& (\overline{v_\phi})^2=\frac2{\sin^2i}[<\widetilde{v}_\|^2>]_{\rm obs}~~; \\
\label{eq:sgo2}
&& \sigma^2=\frac{3-\delta}{1-\delta\cos^2i}[<\sigma_\|^2>]_{\rm obs}~~;
\end{lefteqnarray}
in terms of observed mean rotational velocity and
velocity dispersion.

The intrinsic rotation parameter, by use of
Eqs.\,(\ref{eq:sge2}), (\ref{eq:vfo2}),
(\ref{eq:sgo2}) and (\ref{eq:Erc}), reads:
\begin{equation}
\label{eq:chivo2}
(\chi_v^2)_{\rm int}=\frac{(\overline{v_\phi})^2}{\sigma^2}=\frac2{\sin^2i}
\frac{1-\delta\cos^2i}{3-\delta}\frac{[<\widetilde{v}_\|^2>]_{\rm obs}}
{[<\sigma_\|^2>]_{\rm obs}}~~;
\end{equation}
and sample objects may be represented on the
$({\sf O}\hat{e}\chi_v^2)$ plane using
Eqs.\,(\ref{eq:elin}) and (\ref{eq:chivo2}).

The dimensionless energy:
\begin{equation}
\label{eq:kag}
\kappa=\frac{\sigma_{11}^2a_1}{GM}~~;
\end{equation}
may be considered as the extent to which
gravitation is balanced by centrifugal
forces at the top major axis, $a_1$, as
$-F_{\rm C}/F_{\rm G}\approx\kappa$.   With
regard to [M$_{10}$ kpc Gyr] units,
M$_{10}=10^{10}$m$_\odot$, the constant
of gravitation is $G=4.493\,10^4\,{\rm
M}_{10}^{-1}\,{\rm kpc}^3\,{\rm Gyr}^{-2}$ and
assuming as typical values $M=10^2\,{\rm
M}_{10}$, $a_1=10\,{\rm kpc}$, $\sigma_
{11}^2=4.493\,10^4\,{\rm kpc}\,{\rm Gyr}^
{-1}$, Eq.\,(\ref{eq:kag}) yields $\kappa
=0.1$ as expected for elliptical galaxies.
On the other hand, observed quantities
are related to the effective radius, $R_e$,
instead of the major semiaxis, $a_1$,
and the dimensionless energy has to be
approximated as:
\begin{equation}
\label{eq:kae}
\kappa_{\rm edo}=\frac{[<\sigma_\|^2>]_{\rm edo}R_e}{GM_e}~~;
\end{equation}
in connection with edge-on configurations.   

For the sample of elliptical galaxies under
consideration $(N=16)$, the values of the
following parameters are listed in Tab.\,\ref{t:para}:
\begin{table}
\caption{Parameters calculated from the data 
related to a sample $(N=16)$ of
elliptical galaxies, extracted from larger samples
of early-type galaxies investigated within the
SAURON project (S\,IV, $N=25$; S\,X, $N=48$).
Values of dimensional quantities are expressed in
[M$_{10}$ kpc Gyr] units (1 kpc/Gyr=0.978\,46 km/s;
1 km/s=1.022\,01 kpc/Gyr), and angles in degrees.
Column captions:
(1) NGC number; (2) effective (half-light) radius,
$R_e$ (CV08); (3) intrinsic mean equatorial
tangential velocity component, $\overline{v_\phi}$,
Eq.\,(\ref{eq:vfo2}); (4) intrinsic velocity
dispersion, $\sigma$, Eq.\,(\ref{eq:sgo2});
(5) galaxy stellar mass within the effective radius,
$M_e=M(R_e)=M/2$, under the assumption that luminosity
traces the mass (CV08); (6) dimensionless energy,
$\kappa_{\rm edo}$, Eq.\,(\ref{eq:kae});
%=[<\sigma_\|^2>]_{\rm edo}R_e/(GM_e)$;
(7) inclination angle, $i$, of the best fitting
two-integral Jeans model (S\,IV); (8) anisotropy
parameter, $\delta$, determined from the solution
of the dynamic models, supposed to be axisymmetric
(S\,X); (9) intrinsic ellipticity, $\hat{e}$,
deduced from the computed inclination, Eq.\,(\ref{eq:elin}),
under the
assumption of axisymmetric configurations 
(S\,X); (10) sample object rotation parameter,
$(\chi_v^2)_{\rm int}$, Eq.\,(\ref{eq:chivo2}); 
(11) adjoint configuration rotation parameter,
$\chi_v^2$, Eq.\,(\ref{eq:cii2}); 
(12) kinematic classification, where F and S
denote fast and slow rotators, respectively
(S\,X).   For the original data refer to the parent
papers (S\,IV; S\,X).  For further details refer,
in addition, to earlier attempts (B05;
CV08; C09).}
\label{t:para}
\begin{center}
\begin{tabular}{|c|c|c|c|c|c|c|c|c|c|c|c|} \hline
NGC & $R_e$ & $\overline{v_\phi}$ & $\sigma$ & $M_e$ & $\kappa_{\rm edo}$ &
$i$ & $\delta$ & $\hat{e}$ & $(\chi_v^2)_{\rm int}$ & $\chi_v^2$ & KC
\\
%& (kpc) & (kpc/Gyr) & (kpc/Gyr) & (M$_{10}$) & & ($^\circ$) 
%& & & & & \\
\hline
0821 & 4.43 & 069 & 311 & 10.26 & 0.33 & 90 & 0.20 & 0.40 & 0.05 & 0.35 & F \\
2974 & 2.43 & 219 & 317 & 07.61 & 0.26 & 57 & 0.24 & 0.62 & 0.47 & 0.46 & F \\
3377 & 2.01 & 082 & 198 & 02.35 & 0.27 & 90 & 0.25 & 0.46 & 0.17 & 0.38 & F \\
3379 & 2.09 & 040 & 349 & 08.80 & 0.22 & 90 & 0.03 & 0.08 & 0.01 & 0.05 & F \\
3608 & 4.43 & 012 & 310 & 09.77 & 0.34 & 90 & 0.13 & 0.18 & 0.00 & 0.12 & S \\
4278 & 2.43 & 090 & 410 & 09.64 & 0.33 & 45 & 0.18 & 0.26 & 0.05 & 0.19 & F \\
4374 & 6.15 & 010 & 492 & 36.35 & 0.31 & 90 & 0.08 & 0.15 & 0.00 & 0.09 & S \\
4458 & 2.19 & 014 & 146 & 01.50 & 0.24 & 90 & 0.09 & 0.12 & 0.01 & 0.07 & S \\
4473 & 2.00 & 062 & 318 & 07.86 & 0.22 & 73 & 0.34 & 0.46 & 0.04 & 0.38 & F \\
4486 & 7.96 & 010 & 542 & 45.97 & 0.38 & 90 & 0.00 & 0.04 & 0.00 & 0.02 & S \\
4552 & 2.32 & 019 & 453 & 12.62 & 0.28 & 90 & 0.02 & 0.04 & 0.00 & 0.02 & S \\
4621 & 3.97 & 075 & 355 & 18.80 & 0.21 & 90 & 0.18 & 0.34 & 0.04 & 0.27 & F \\
4660 & 0.67 & 122 & 279 & 02.11 & 0.20 & 70 & 0.30 & 0.53 & 0.19 & 0.40 & F \\
5813 & 7.90 & 046 & 389 & 28.89 & 0.32 & 90 & 0.08 & 0.15 & 0.01 & 0.09 & S \\
5845 & 0.56 & 117 & 390 & 03.02 & 0.22 & 90 & 0.15 & 0.35 & 0.09 & 0.29 & F \\
5846 & 9.51 & 010 & 424 & 37.19 & 0.34 & 90 & 0.01 & 0.07 & 0.00 & 0.04 & S \\
\hline
\end{tabular}
\end{center}
\end{table}
the effective (half-light) radius,
$R_e$ (CV08); the intrinsic mean equatorial
tangential velocity component, $\overline{v_\phi}$,
Eq.\,(\ref{eq:vfo2}); the intrinsic velocity
dispersion, $\sigma$, Eq.\,(\ref{eq:sgo2});
the galaxy stellar mass within the effective radius,
$M_e=M(R_e)=M/2$, under the assumption that luminosity
traces the mass (CV08); the dimensionless energy,
$\kappa_{\rm edo}$, Eq.\,(\ref{eq:kae});
%=[<\sigma_\|^2>]_{\rm edo}R_e/(GM_e)$;
the inclination angle, $i$, of the best fitting
two-integral Jeans model (S\,IV); the anisotropy
parameter, $\delta$, determined from the solution
of the dynamic models, supposed to be axisymmetric
(S\,X); the intrinsic ellipticity, $\hat{e}$,
deduced from the computed inclination, Eq.\,(\ref
{eq:elin}), under the
assumption of axisymmetric configurations 
(S\,X); the sample object rotation parameter,
$(\chi_v^2)_{\rm int}$, Eq.\,(\ref{eq:chivo2}); 
the adjoint configuration rotation parameter,
$\chi_v^2$, Eq.\,(\ref{eq:cii2}); together with
the kinematic classification, where F and S
denote fast and slow rotators, respectively
(S\,X).   For the original data refer to the parent
papers (S\,IV; S\,X).  For further details refer,
in addition, to earlier attempts (B05;
CV08; C09).   It can also be seen that
$\kappa_{\rm edo}=0.20$-0.34 for the whole sample, which
implies gravitation is not balanced by centrifugal
force in elliptical galaxies, as expected.

\begin{figure*}[t]
\begin{center}
\includegraphics[scale=0.8]{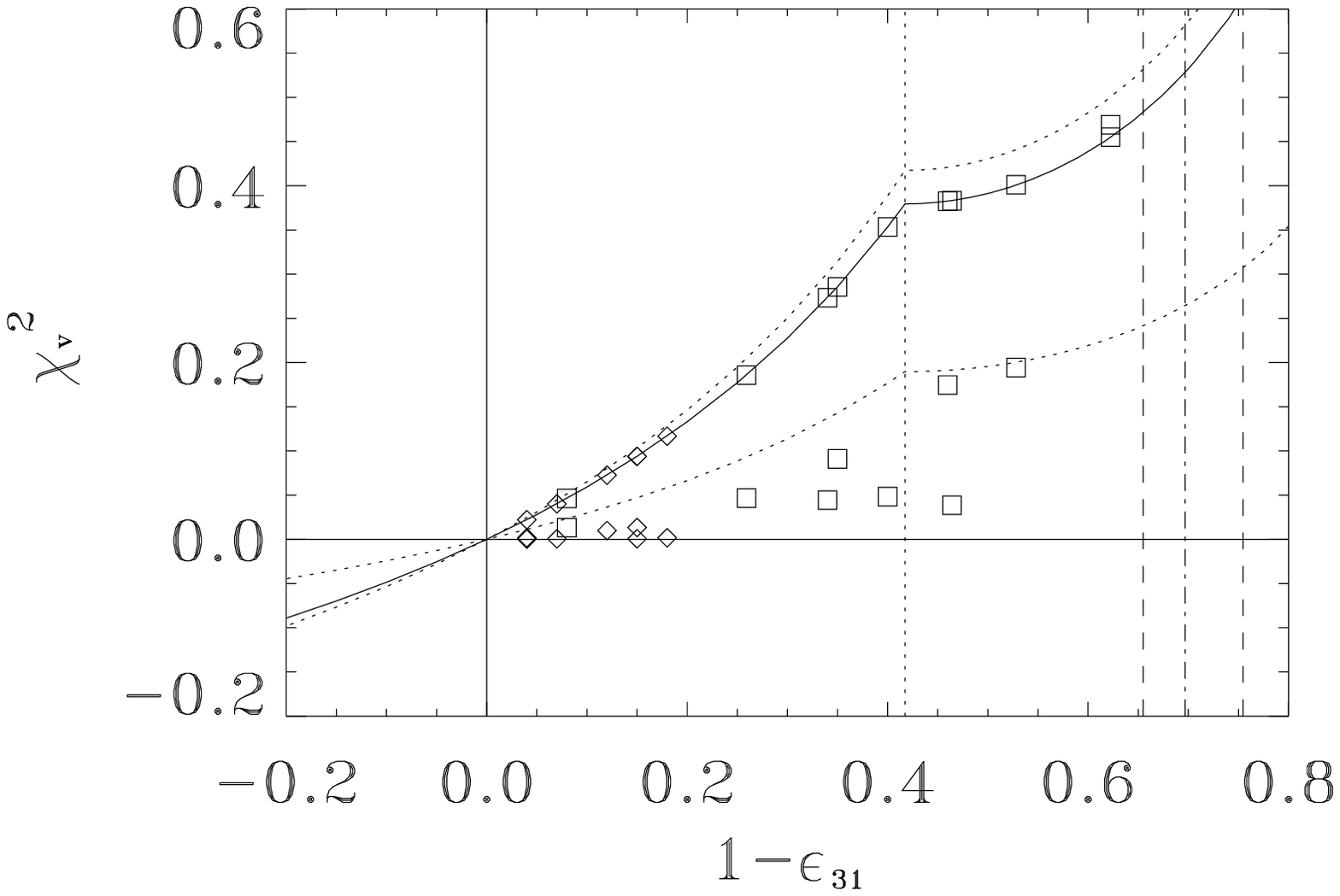}
\caption{Position of elliptical galaxies listed in
Tab.\,\ref{t:para}, on the $({\sf O}\hat{e}\chi_v^2)$ plane,
denoted as squares and diamonds for fast and slow rotators,
respectively.   Curves are taken from Fig.\,\ref{f:seq},
related to equilibrium figures ($\zeta=1$, full line) and
nonequilibrium figures ($\zeta=1.1$, up on the first quadrant;
$\zeta=0.5$, down on the first quadrant;
dotted lines) with isotropic stress tensor.   Caption of
vertical lines as in Fig.\,\ref{f:seq}.   The adjoint
configurations with isotropic stress tensor
and equal shape, are also positioned
on the sequence of equilibrium figures using the same
symbols, as the result of a vertical shift.   A similar
procedure holds, in general, for an assigned sequence
of nonequilibrium figures with isotropic stress tensor.}
\label{f:ellr}
\end{center}
\end{figure*}

The position of fast and slow rotators on
the $({\sf O}\hat{e}\chi_v^2)$ plane, is
shown in Fig.\,\ref{f:ellr} as squares and
diamonds, respectively.
Three curves are also reproduced from Fig.\,\ref{f:seq},
namely equilibrium figures ($\zeta=1$, full line) and
nonequilibrium figures ($\zeta=1.1$, up on the first
quadrant; $\zeta=0.5$, down on the first quadrant;
dotted lines) with isotropic stress tensor.
The caption of the vertical lines is the same as in
Fig.\,\ref{f:seq}.   The adjoint configurations with
isotropic stress tensor
and equal shape, are also positioned on the sequence
of equilibrium figures using the same symbols, as the
result of a vertical shift.   A similar procedure
holds, in general, for an assigned sequence of
nonequilibrium figures with isotropic stress tensor.
An inspection of Fig.\,\ref{f:ellr} shows the
following features.

All sample objects lie below the lower sequence
($\zeta=0.5$), with the exception of the most
flattened galaxy (NGC 2974), which is placed just
above the middle sequence ($\zeta=1$).

In general, slow rotators (diamonds) are placed
on the left with respect to fast rotators (squares)
with a single exception (NGC 3379) which, on the
other hand, shows a nearly isotropic stress tensor
$(\delta=0.03)$, a nearly spherical shape $(\hat
{e}=0.04)$, and a velocity ratio, $\{[<\widetilde
{v}_\|^2>]_{\rm obs}/[<\sigma_\|^2>]_{\rm obs}\}^
{1/2}=0.14$, equal to the maximum value found in
slow rotators (NGC 5813).   A similar trend is
exhibited by adjoint configurations with isotropic
stress tensor, which
are vertically shifted on the sequence of equilibrium
figures, full line in Fig.\,\ref{f:ellr}.

If slow rotators (including NGC 3379) are assumed
to be nonrotating $(\overline{v_\phi}=0)$ within
the observational errors and elongated $(\hat{e}<
0)$ due to an anisotropic stress tensor ($\sigma_{11}
=\sigma_{22}<\sigma_{33}$ or $\delta<0$), the
related adjoint configurations with isotropic stress
tensor are elongated due to imaginary rotation,
conformly to Eq.\,(\ref{eq:cii2}).   More specifically,
the major axis coincides with the polar axis and the
meridional axis ratio exceeds unity, yielding a
negative ellipticity, $\hat{e}=1-\epsilon_{31}<0$,
which produces, in turn, a negative anisotropy parameter,
$\delta=1-\sigma_{33}^2/\sigma_{11}^2<0$, by
generalizing Eq.\,(\ref{eq:deltaa}) to the case
under discussion, $\sigma_{11}=\sigma_{22}<\sigma_{33}$.

In this view, let $\hat{e}_{\rm fla}$, $\delta_
{\rm fla}$, be the intrinsic ellipticity from
Eq.\,(\ref{eq:elin}) and the anisotropy parameter
determined by comparison with the dynamic models,
for axisymmetric flattened configurations (S\,X),
and $\hat{e}_{\rm elo}$, $\delta_{\rm elo}$,
their counterparts related to equal values of
intrinsic axes and stress tensor components,
but with the major axis related to the symmetry
axis instead of the equatorial plane.   Accordingly,
the following changes must be performed: $\epsilon_
{31}\to\epsilon_{31}^{-1}$; $\sigma_{33}\to\sigma_{11}$;
$\sigma_{11}\to\sigma_{33}$; which, using Eqs.\,(\ref
{seq:delta}) and (\ref{eq:elin}), yields:
\begin{lefteqnarray}
\label{eq:elel}
&& \hat{e}_{\rm elo}=1-\epsilon_{31}^{-1}=1-(1-\hat{e}_{\rm fla})^{-1}=
\frac{-\hat{e}_{\rm fla}}{1-\hat{e}_{\rm fla}}~~; \\
\label{eq:deel}
&& \delta_{\rm elo}=1-\left(\frac{\sigma_{33}}{\sigma_{11}}\right)^{-2}=
1-(1-\delta_{\rm fla})^{-1}=\frac{-\delta_{\rm fla}}{1-\delta_{\rm fla}}~~;
\end{lefteqnarray}
where the values related to flattened configurations are
listed in Tab.\,\ref{t:para} and again in Tab.\,\ref{t:pasr}
for slow rotators (NGC 3379 included), together with
their counterparts related to elongated configurations
and, in both cases, the values of the rotation parameter,
$\chi_v$, for adjoint configuration with isotropic stress
tensor.
\begin{table}
\caption{The intrinsic ellipticity, $\hat{e}$,
the anisotropy parameter, $\delta$, and the
square root of the absolute value of the
rotation parameter, $\vert\chi_v^2\vert^{1/2}$,
for slow rotators
(NGC 3379 included) supposed to be flattened (fla)
as in Tab.\,\ref{t:para}, or elongated (elo) with
equal major axis, according to Eqs.\,(\ref{eq:elel})
and (\ref{eq:deel}).   The square root of the
absolute value of the
rotation parameter has been listed for better
appreciating the difference between flattened
and elongated configurations.}
\label{t:pasr}
\begin{center}
\begin{tabular}{|c|c|c|c|c|c|c|} \hline
NGC & $\hat{e}_{\rm fla}$ & $\delta_{\rm fla}$ & $\vert(\chi_v^2)_{\rm fla}
\vert^{1/2}$ & $-\hat{e}_{\rm elo}$ & $-\delta_{\rm elo}$ & $\vert(\chi_v^2)_
{\rm elo}\vert^{1/2}$ \\
\hline
3379 & 0.08 & 0.03 & 0.22 & 0.09 & 0.03 & 0.21 \\
3608 & 0.18 & 0.13 & 0.34 & 0.22 & 0.15 & 0.31 \\
4374 & 0.15 & 0.08 & 0.31 & 0.18 & 0.09 & 0.28 \\
4458 & 0.12 & 0.09 & 0.27 & 0.14 & 0.10 & 0.25 \\
4486 & 0.04 & 0.00 & 0.15 & 0.04 & 0.00 & 0.15 \\
4552 & 0.04 & 0.02 & 0.15 & 0.04 & 0.02 & 0.15 \\
5813 & 0.15 & 0.08 & 0.31 & 0.18 & 0.09 & 0.28 \\
5846 & 0.07 & 0.01 & 0.20 & 0.08 & 0.01 & 0.19 \\
\hline
\end{tabular}
\end{center}
\end{table}
The square root of the
rotation parameter has been listed for better
appreciating the difference between flattened
and elongated configurations.

\begin{figure*}[t]
\begin{center}
\includegraphics[scale=0.8]{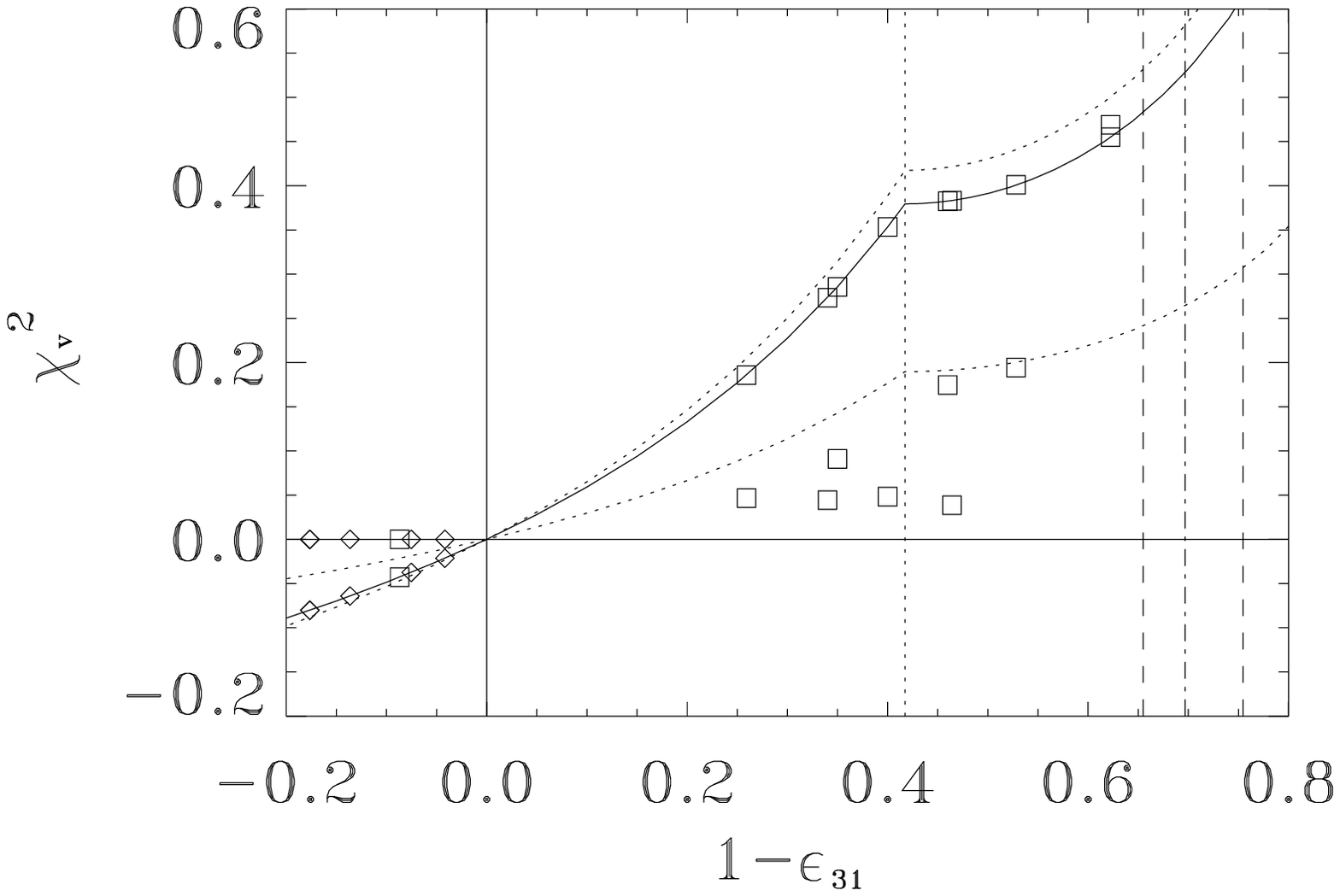}
\caption{Same as in Fig.\,\ref{f:ellr}, but under the
assumption that slow rotators (NGC 3379 included)
are nonrotating within the observational errors and
elongated due to an anisotropic stress tensor.
For further details refer to the text.   Captions
as in Fig.\,\ref{f:ellr}.}
\label{f:esre}
\end{center}
\end{figure*}

The position of slow rotators (NGC 3379 included)
on the $({\sf O}\hat{e}\chi_v^2)$ plane in the
case under discussion, is indicated in Fig.\,\ref
{f:esre} as diamonds and a single square, while
nothing changes for the remaining
fast rotators (squares) and captions, with respect
to Fig.\,\ref{f:ellr}.

\section{Comparison with earlier attempts}
\label{chCB}

For equilibrium figures $(\zeta=1)$ viewed
edge-on $(i=90^\circ)$, the combination of
Eqs.\,(\ref{seq:Epot}), (\ref{seq:chi2}),
(\ref{eq:delzi}), and (\ref{eq:chivo2})
yields:
\begin{equation}
\label{eq:cheo}
(\chi_v^2)_{\rm C}=\frac{[<\widetilde{v}_\|^2>]_{\rm edo}}{[<\sigma_\|^2>]_
{\rm edo}}=(1-\delta)\frac{(B_{\rm sel})_{11}}{(B_{\rm sel})_{33}}-1~~;
\end{equation}
which is formally coincident with classical
results, where $<\widetilde{v}_\|^2>$ and
$<\sigma_\|^2>$ are
intended as a mass-averaged, one-dimension
squared rotation velocity and related
variance, respectively (e.g., B05).

If, on the other hand, $<\widetilde{v}_\|^2>$
and $<\sigma_\|^2>$ are conceived as sky-averages,
according to the current notation,
and the kinetic energy is decomposed into
contributions from ordered and random motions,
the right-hand
side of Eq.\,(\ref{eq:cheo}) has to be divided by
a correction factor.   The result is (B05):
\begin{equation}
\label{eq:cheoB}
(\chi_v^2)_{\rm B}=\frac{[<\widetilde{v}_\|^2>]_{\rm edo}}{[<\sigma_\|^2>]_
{\rm edo}}=\frac{(1-\delta)(B_{\rm sel})_{11}/(B_{\rm sel})_{33}-1}
{\psi(1-\delta)(B_{\rm sel})_{11}/(B_{\rm sel})_{33}+1}~~;
\end{equation}
where $\psi$ is an integral which depends on
the density profile and the rotation curve of
the matter distribution.

In the present investigation,
Eq.\,(\ref{eq:cheo}) holds
provided the kinetic energy is
decomposed into contributions
from cylindrical rotation and
residual motions, which implies
a different anisotropy parameter
with respect to Eq.\,(\ref{eq:cheoB}).
The left-hand sides of
Eqs.\,(\ref{eq:cheo}) and (\ref{eq:cheoB}):
\begin{equation}
\label{eq:chedo}
(\chi_v^2)_{\rm edo}=\frac{[<\widetilde{v}_\|^2>]_{\rm edo}}{[<\sigma_\|^2>]_
{\rm edo}}~~;
\end{equation}
must necessarily coincide, if deduced
from observations, which is true also
for the shape factors.   Accordingly,
the combination of Eqs.\,(\ref{eq:cheo})
and (\ref{eq:cheoB}) yields:
\begin{equation}
\label{eq:deCB}
(1-\delta_{\rm C})\frac{(B_{\rm sel})_{11}}{(B_{\rm sel})_{33}}-1=
\frac{(1-\delta_{\rm B})(B_{\rm sel})_{11}/(B_{\rm sel})_{33}-1}
{\psi(1-\delta_{\rm B})(B_{\rm sel})_{11}/(B_{\rm sel})_{33}+1}~~;
\end{equation}
where the indices, C and B, denote
anisotropy parameters related to
the current and earlier (B05)
attempt, respectively.

The values of the anisotropy parameter
for sample objects, listed in Tab.\,\ref{t:para},
are taken from an earlier attempt (S\,X),
where they have been determined in accordance
with Eq.\,(\ref{eq:cheoB}).    For this
reason, a comparison is needed between
values of the rotation parameter, $(\chi_v^2)_
{\rm C}$ and $(\chi_v^2)_{\rm B}$, defined by
Eqs.\,(\ref{eq:cheo}) and (\ref{eq:cheoB}),
respectively.   To this aim, the rotation
parameter ratio:
\begin{equation}
\label{eq:chCB}
\chi_{\rm CB}^2=\frac{(\chi_v^2)_{\rm C}}{(\chi_v^2)_{\rm B}}=
\psi(1-\delta_{\rm B})\frac{(B_{\rm sel})_{11}}{(B_{\rm sel})_{33}}+1~~;
\qquad0<\epsilon_{31}<1~~;
\end{equation}
shall be considered, which is maximum for
isotropic stress tensor, $\delta=0$.   In
the spherical limit, $\epsilon_{31}=1$,
$(B_{\rm sel})_{11}=(B_{\rm sel})_{33}$,
which implies $\delta=0$, and Eq.\,(\ref{eq:chCB})
reduces to:
\begin{equation}
\label{eq:chCBs}
\chi_{\rm CB}^2=\psi+1~~;\qquad\epsilon_{31}=1~~;
\end{equation}
in the flat limit, $\epsilon_{31}=0$,
$(B_{\rm sel})_{11}=\pi/2$, $(B_{\rm sel})_{33}=0$,
which implies $\chi_{\rm CB}^2\to+\infty$
unless $\delta=1$.   In dealing with
elliptical galaxies, $\epsilon_{31}\ge0.3$,
the flat limit is never attained.
For realistic density profiles,
numerical computations yield
$\psi\approx0.131$ in presence
of flat rotation curves, $\overline
{v_\phi}={\rm const}$ (B05), and
$\psi\approx0.15$ using Jeans models
(S\,X).

The results are listed in Tab.\,\ref{t:chCB},
where the values of the following parameters
are calculated or deduced from Tab.\,\ref{t:para}
for sample objects:
\begin{table}
\caption{Comparison between intrinsic rotation
parameters, $(\chi_v^2)_{\rm C}$ and $(\chi_v^2)_{\rm B}$,
calculated for sample objects listed in Tab.\,\ref{t:para},
using Eqs.\,(\ref{eq:cheo}) and (\ref{eq:cheoB}),
respectively, for isotropic stress tensor $(\delta=0)$
and $\psi=0.15$ (S\,X).   Also listed are values of the
ratios, $\chi_{\rm CB}^2=(\chi_v^2)_{\rm C}/(\chi_v^2)_{\rm B}$ and
$p=1/(\sqrt{2}\chi_{\rm CB})$.   The remaining parameters are taken
or deduced from Tab.\,\ref{t:para}.  For meridional axis ratios
below the bifurcation point, $\epsilon_{31}<(\epsilon_{31})_{\rm bif}=
0.582\,724$, the shape factors are determined for triaxial instead of
axisymmetric configurations, where the minor equatorial axis is assumed 
to be parallel to the line of sight.   Column captions:
(1) NGC number; (2) intrinsic meridional axis ratio, $\epsilon_{31}=
1-\hat{e}$, Eq.\,(\ref{eq:elin});
(3) shape factor ratio, $b_{23}=(B_{\rm sel})_{22}/(B_{\rm sel})_{33}$,
Eq.\,(\ref{eq:Epotb});
(4) inclination angle, $i$, of the best fitting
two-integral Jeans model (S\,IV); (5) anisotropy
parameter, $\delta$, determined from the solution
of the dynamic models, supposed to be axisymmetric
(S\,X); (6) sample object rotation parameter,
$(\chi_v^2)_{\rm edo}=[<\widetilde{v}_\|^2>]_{\rm edo}/
[<\sigma_\|^2>]_{\rm edo}$, Eqs.\,(\ref{eq:vrin})
and (\ref{eq:sgin});
(7) adjoint configuration rotation parameter,
$(\chi_v^2)_{\rm C}$, Eq.\,(\ref{eq:cheo}); 
(8) adjoint configuration rotation parameter,
$(\chi_v^2)_{\rm B}$, Eq.\,(\ref{eq:cheoB}); 
(9) rotation parameter ratio, $\chi_{\rm CB}^2$,
Eq.\,(\ref{eq:chCB});
(10) ratio, $p=1/(\sqrt{2}\chi_{\rm CB})$;
(11) kinematic classification, where F and S
denote fast and slow rotators, respectively
(S\,X).}
\label{t:chCB}
\begin{center}
\begin{tabular}{|c|c|c|c|c|c|c|c|c|c|c|} \hline
NGC & $\epsilon_{31}$ & $b_{23}$ & $i$ & $\delta$ & $(\chi_v^2)_{\rm edo}$ & 
$(\chi_v^2)_{\rm C}$ & $(\chi_v^2)_{\rm B}$ & $\chi^2_{\rm CB}$ &
$p$ & KC
\\
%& (kpc) & (kpc/Gyr) & (kpc/Gyr) & (M$_{10}$) & & ($^\circ$) 
%& & & & & \\
\hline
0821 & 0.60 & 1.53 & 90 & 0.20 & 0.07 & 0.35 & 0.29 & 1.23 & 0.64 & F \\
2974 & 0.38 & 2.29 & 57 & 0.24 & 0.65 & 0.46 & 0.38 & 1.19 & 0.65 & F \\
3377 & 0.54 & 1.68 & 90 & 0.25 & 0.24 & 0.38 & 0.31 & 1.22 & 0.64 & F \\
3379 & 0.92 & 1.07 & 90 & 0.03 & 0.02 & 0.05 & 0.04 & 1.16 & 0.66 & F \\
3608 & 0.82 & 1.18 & 90 & 0.13 & 0.00 & 0.12 & 0.10 & 1.18 & 0.65 & S \\
4278 & 0.74 & 1.28 & 45 & 0.18 & 0.07 & 0.19 & 0.16 & 1.19 & 0.65 & F \\
4374 & 0.85 & 1.14 & 90 & 0.08 & 0.00 & 0.09 & 0.08 & 1.17 & 0.65 & S \\
4458 & 0.88 & 1.11 & 90 & 0.09 & 0.01 & 0.07 & 0.06 & 1.17 & 0.65 & S \\
4473 & 0.54 & 1.68 & 73 & 0.34 & 0.05 & 0.38 & 0.31 & 1.22 & 0.64 & F \\
4486 & 0.96 & 1.03 & 90 & 0.00 & 0.00 & 0.02 & 0.02 & 1.15 & 0.66 & S \\
4552 & 0.96 & 1.03 & 90 & 0.02 & 0.00 & 0.02 & 0.02 & 1.15 & 0.66 & S \\
4621 & 0.66 & 1.41 & 90 & 0.18 & 0.06 & 0.27 & 0.23 & 1.21 & 0.64 & F \\
4660 & 0.47 & 1.89 & 70 & 0.30 & 0.26 & 0.40 & 0.33 & 1.21 & 0.64 & F \\
5813 & 0.85 & 1.14 & 90 & 0.08 & 0.02 & 0.09 & 0.08 & 1.17 & 0.65 & S \\
5845 & 0.65 & 1.43 & 90 & 0.15 & 0.13 & 0.29 & 0.24 & 1.21 & 0.64 & F \\
5846 & 0.93 & 1.06 & 90 & 0.01 & 0.00 & 0.04 & 0.03 & 1.16 & 0.66 & S \\
\hline
\end{tabular}
\end{center}
\end{table}
the intrinsic meridional axis ratio,
$\epsilon_{31}=1-\hat{e}$, Eq.\,(\ref{eq:elin});
the shape factor ratio, $b_{23}=(B_{\rm sel})_
{22}/(B_{\rm sel})_{33}$, Eq.\,(\ref{eq:Epotb});
the inclination angle, $i$, of the best fitting
two-integral Jeans model (S\,IV); the anisotropy
parameter, $\delta$, determined from the solution
of the dynamic models, supposed to be axisymmetric
(S\,X); the sample object rotation parameter,
$(\chi_v^2)_{\rm edo}=[<\widetilde{v}_\|^2>]_{\rm edo}/
[<\sigma_\|^2>]_{\rm edo}$, Eqs.\,(\ref{eq:vrin})
and (\ref{eq:sgin});
the adjoint configuration rotation parameter,
$(\chi_v^2)_{\rm C}$, Eq.\,(\ref{eq:cheo}); 
the adjoint configuration rotation parameter,
$(\chi_v^2)_{\rm B}$, Eq.\,(\ref{eq:cheoB}); 
the rotation parameter ratio, $\chi_{\rm CB}^2$,
Eq.\,(\ref{eq:chCB});
the ratio, $p=1/(\sqrt{2}\chi_{\rm CB})$; together
with the kinematic classification, where F and S
denote fast and slow rotators, respectively
(S\,X).   For sample objects with meridional
axis ratio below the bifurcation point, 
$\epsilon_{31}<(\epsilon_{31})_{\rm bif}=
0.582\,724$, the shape factors are determined for 
triaxial instead of axisymmetric configurations,
where the minor equatorial axis is assumed to be
parallel to the line of sight.

An inspection of Tab.\,\ref{t:chCB}
shows that $\chi_{\rm CB}^2=$1.15-1.23
and $1/(\sqrt{2}\chi_{\rm CB})=$
0.64-0.66.   Curiously, the last
range is consistent with the squared
coefficient, $2\times0.57^2$, of
the best fitting relation between
rotation parameters derived from
integral-field and long-slit stellar 
kinematics (S\,X), the second expressed
in terms of one-dimension peak velocity.

\section{Discussion}
\label{disc}

Though the generalization of the
ellipticity-rotation plane,
$({\sf O}\hat{e}\chi_v^2)$,
takes nonequilibrium figures into
consideration, still galaxies may
be thought of as fully virialized
$(\zeta=1)$ unless a major merger
is going on.   On the other hand,
cluster of galaxies are presently
assembling and radial motions dominate
at large clustercentric distances,
which implies dynamical evolution
$(\zeta\ne1)$.   For this reason,
only the sequence of equilibrium
figures has been considered in
dealing with elliptical galaxies.
The rotation parameter, $\chi_v^2$,
depends on the intrinsic squared
mean equatorial tangential velocity
component and velocity component
dispersions,
according to Eq.\,(\ref{seq:chi2}).

A different rotation parameter has been
defined in an earlier attempt (B05),
which depends on the sky-averaged
mean streaming velocity parallel to the
line of sight, the sky-averaged streaming
velocity (parallel to the line of sight)
dispersion, and the sky-averaged component
velocity (parallel to the line of sight)
dispersion related to nonstreaming motions.

The inclination angle, $i$, and the
anisotropy parameter, $\delta$, have
been determined for sample objects by
comparison with the dynamic models,
under the assumption of axisymmetric
configurations $(a_1=a_2)$, axisymmetric
stress tensor $(\sigma_{11}=\sigma_{22})$,
and flattened shapes $(a_1>a_3$, $\sigma_{11}
>\sigma_{33})$, as outlined in the parent
paper (S\,X).   On the other hand, it is
shown in Fig.\,\ref{f:ellr} that four
sample objects are predicted to be triaxial
as the ellipticity value related to the
bifurcation point $(\hat{e}_{\rm bif}=
0.417\,276)$, marked by the vertical dotted
line, is exceeded, and the most flattened
configurations could be barlike (e.g.,
$\epsilon_{21}=0.432\,232$ for $\epsilon_{31}
=0.345\,069$).   Then some caution must be
adopted in the interpretation of the trixiality,
expecially for the most aspherical sample object
(NGC 2974).   To this respect, it is worth
remembering that galaxies are embedded in
dark (nonbaryonic) matter haloes, according
to current cosmological scenarios.

The presence of a massive, embedding
subsystem stabilizes the inner spheroid
and shifts the bifurcation point (from
axisymmetric to triaxial configurations)
towards increasing ellipticities (Durisen
1978; Pacheco et al. 1986).   In the
special case of homogeneous subsystems,
the bifurcation point is attained at
$\hat{e}\approx0.7$ for comparable masses
within the volume of the inner spheroid
(Caimmi 1996a), which can be considered
as an upper limit.   In fact, by comparison
with the dynamic models, fast rotators
appear to be axisymmetric while,
paradoxically, slow rotators exhibit a
moderate triaxiality (S\,X).   Using
triaxial dynamic models could provide
better understanding on this point.
An inspection of Fig.\,\ref{f:ellr}
and Tab.\,\ref{t:para} shows that,
for slow rotators (including NGC 3379)
$0\le\hat{e}<0.2$; $0\le\delta<0.15$;
$0\le\chi_v^2<0.15$; and for fast rotators
(excluding NGC 3379) $0.2\le\hat{e}<0.65$;
$0.15\le\delta<0.35$; $0.15\le\chi_v^2<0.5$;
which could be an alternative kinematic
classification with respect to earlier
attempts (e.g., Emsellem et al. 2007).
Richer samples should be dealt with to
provide more conclusive evidence on
this point.

If slow rotators (including NGC 3379)
are nonrotating within the observational
errors, and their shapes are elongated
due to negative anisotropy parameters,
$\delta<0$ i.e. $\sigma_{11}=\sigma_{22}
<\sigma_{33}$, a different kinematic
classification is shown in Fig.\,\ref
{f:esre} and Tab.\,\ref{t:pasr}.
With regard to adjoint configurations
with isotropic stress tensor, fast and
slow rotators are affected by real and
imaginary systematic rotation, respectively.
It is, of course, a limiting situation,
in the sense that no zone of avoidance
is expected on the sequence
of adjoint configurations.   If nonrotating
elongated configurations really exist,
slow rotators should be divided into
two subclasses, namely (1) flattened
in real rotatin, and (2) elongated in
imaginary rotation.   In the latter
alternative,
elongated instead of flattened dynamic
models should be used for determining
the inclination angle and the anisotropy
parameter by comparison with the data
(S\,IV; S\,X), to gain consistency and
to test the above interpretation.

Given a spherical galaxy with isotropic
stress tensor, rotation (around a
selected axis) kinetic energy may
be added in infinite ways between
two limiting situations, namely (a)
systematic rotation, where circular
motions are either clockwise or counterclockwise,
and (b) random rotation, where the mean
circular velocity is null.   In any case,
all the observables remain unchanged with
the exception of the angular momentum
and the stress tensor.   More specifically,
any system may be related to its adjoint
configuration with isotropic stress tensor,
without loss of generality, for ellipticity
values above a threshold (Appendix \ref{a:rost}).

In this view, the elliptical side of the
Hubble sequence may be interpreted as a
sequence of equilibrium (adjoint) figures
where the ellipticity, $\hat{e}$, is
increasing with the rotation parameter,
$\chi_v^2$.   The original (elliptical
side of the) Hubble sequence may be
generalized on two respects, namely
(1) from equilibrium $(\zeta=1)$ to
nonequilibrium $(\zeta\ne1)$ figures,
and (2) from real $(\hat{e}\ge0,\,
\chi_v^2\ge0)$ to imaginary $(\hat{e}<0,\,
\chi_v^2<0)$ rotation.   More specifically,
real rotation is related to flattened
configurations and imaginary rotation to
elongated configurations, spinning around
the minor and the major axis, respectively.

The above classification is grounded on
the assumption of homeoidally striated
density profiles, regardless of the
mechanism of formation.   More specifically,
configurations of equal shape related to
different assembling processes belong to
the same class.   In reality, isophotes
in elliptical galaxies may be boxy i.e.
``overelliptic'' or disky i.e. ``underelliptic''
where, in any case, the axis ratio changes
with radius.   This dichotomy, together with
the one related to dwarf ang giant elliptical
galaxies, is interpreted as due to different
past histories (e.g., Kormendy et al. 2009).
On the other hand, a simplified description
in terms of isophotes with constant axis
ratio at all radii, implies a classification
of equilibrium (or nonequilibrium with fixed
virial index) figures which is independent of
the formation and evolution processes.

\section{Conclusion}\label{conc}

The results of earlier investigations
on homeoidally striated MacLaurin
spheroids and Jacobi ellipsoids (CM05;
Caimmi 2006a; C07) are used in the
current attempt for the representation
of nonequilibrium figures on the
ellipticity-velocity plane, and of
figures in imaginary rotation, where
the effect is elongating instead of
flattening, with respect to the
rotation axis.   The key concept
is that the addition of kinetic
energy related to tangential equatorial
velocity components makes distorted
boundaries regardless of what fraction
translates in an increment of angular
momentum, and what fraction in an
increment of stress tensor
equatorial components.
Then any system admits an adjoint
configuration with isotropic stress
tensor, for ellipticity
values above a threshold (Appendix \ref{a:rost}),
and the related sequence can
be considered without loss of generality.
The kinetic energy is decomposed into
contributions from cylindrical rotation,
$E_{\rm rot}=M(\overline{v_\phi})^2/2$,
and tangential equatorial component velocity
dispersion plus residual motions, $E_{\rm res}=M
(\sigma_{\phi\phi}^2+\sigma_{ww}^2+
\sigma_{33}^2)/2$, and the rotation
parameter is defined as $\chi_v^2=
E_{\rm rot}/E_{\rm res}$.

Elliptical galaxies are idealized as
homeoidally striated MacLaurin spheroids
and Jacobi ellipsoids, and their position
on the ellipticity-velocity plane is
inferred from observations related to
a sample (CV08, $N=16$) extracted from
richer samples of early-type galaxies
investigated within the SAURON project
(S\,IV, $N=25$; S\,X, $N=48$).   The
location of model galaxies on the
$({\sf O}\hat{e}\chi_v^2)$ plane is
determined through the following steps
(C09): (i) select SAURON data of interest;
(ii) calculate the parameters appearing
in the virial equations; (iii) make a
correspondence between model galaxies
and sample objects; (iv) represent model
galaxies on the $({\sf O}\hat{e}\chi_v^2)$
plane.

The main results found in the present
investigation may be summarized as
follows.
\begin{description}
\item[(1)]
Sequences of homeoidally striated
MacLaurin spheroids and Jacobi ellipsoids
with isotropic stress tensor are defined
and plotted on the $({\sf O}\hat{e}\chi_v^2)$
plane, without loss of generality: for
any system with anisotropic stress tensor
an adjoint configuration exists, with
isotropic stress tensor and remaining
parameters unchanged with the exception
of the angular momentum, for ellipticity
values above a threshold.
\item[(2)]
Sequences of homeoidally striated
MacLaurin spheroids and Jacobi ellipsoids
are generalized to nonequilibrium figures.
\item[(3)]
Sequences of homeoidally striated
MacLaurin spheroids and Jacobi ellipsoids
are generalized to imaginary rotation
which acts in elongating, instead of
flattening, with respect to the rotation
axis.
\item[(4)]
An alternative kinematic classification
with respect to earlier attempts (e.g.,
Emsellem et al. 2007) is proposed, where
slow rotators are characterized by low
ellipticities $(0\le\hat{e}<0.2)$, low
anisotropy parameters $(0\le\delta<0.15)$,
and low rotation parameters $(0\le\chi_v^2<
0.15)$, and fast rotators by large ellipticities
$(0.2\le\hat{e}<0.65)$, large anisotropy
parameters $(0.15\le\delta<0.35)$, and
large rotation parameters $(0.15\le\chi_v^2<0.5)$.
Richer samples should be used to test
the validity of the above interpretation.
\item[(5)]
A possible interpretation of slow rotators
as nonrotating and elongated due to a negative
anisotropy parameter, instead of flattened due
to a positive anisotropy parameter, is exploited.
\item[(6)]
The elliptical side of the Hubble morphological
sequence is interpreted as a sequence of equilibrium
(adjoint) figures where the ellipticity is an
increasing function of the rotation parameter.
Accordingly, slow rotators correspond to early
classes (E0-E2 in the oblate limit and E$-$2-E0
in the prolate limit) and fast rotators to late
classes (E3-E6).   In this view, boundaries are
rotationally distorted regardless of what fraction
of tangential equatorial velocity components is
related to angular momentum, and what fraction
to stress tensor equatorial components.
\end{description}

\section{Acknowledgements}
%We are indebted to an anonymous referee for critical
%comments which  improved an
%earlier version of the manuscript.
Thanks are due to T. Valentinuzzi for fruitful discussions.
%The analytical integrations needed in the current paper
%were helped substantially by use of the Mathematica
%package and visiting the internet site:
%``HTTP://INTEGRALS. WOLFRAM.COM/INDEX.CGI''.   This
%is why we are deeply grateful to the Wolfram staff,
%in particular to Daniel Lichtblau, and wish to
%acknowledge all the facilities encountered
%therein.

\appendix
\section*{Appendix}

\section{Rotation and residual kinetic energy}
\label{a:ciro}

With regard to a generic matter distribution, let
$({\sf O}x_1x_2x_3)$ be a reference frame where
the origin coincides with the centre of mass,
and the coordinate axis, $x_3$, coincides with
the rotation axis.   Let $v_\phi(x_1,x_2,x_3,t)$
be the tangential equatorial (with respect to the
rotation axis) velocity component,
related to the generic particle at the point,
${\sf P}(x_1,x_2,x_3)$, at the time, $t$.  The
corresponding angular velocity is defined by
the relation:
\begin{equation}
\label{eq:vfO}
v_\phi(x_1,x_2,x_3,t)=\Omega(x_1,x_2,x_3,t)w~~;\qquad w^2=x_1^2+x_2^2~~;
\end{equation}
which can also be extended to mean values,
$\overline{v_\phi}(x_1,x_2,x_3,t)$,
$\overline{\Omega}(x_1,x_2,x_3,t)$,
whithin an infinitesimal volume element,
$\diff x_1\diff x_2\diff x_3$, placed at
the same point, {\sf P}.   In the following
relations, the time dependence shall be
omitted on the quantities at the left-hand
side, to gain simplicity.

The mass-weighted tangential equatorial
velocity component and squared tangential
equatorial velocity component read:
\begin{lefteqnarray}
\label{eq:vphi}
&& \overline{v_\phi}=\frac1M\int\int\int\overline{v_\phi}(x_1,x_2,x_3,t)
\rho(x_1,x_2,x_3,t)\diff x_1\diff x_2\diff x_3~~; \\
\label{eq:vph2}
&& \overline{(v_\phi^2)}=\frac1M\int\int\int\overline{v_\phi^2}
(x_1,x_2,x_3,t)\rho(x_1,x_2,x_3,t)\diff x_1\diff x_2\diff x_3~~;
\end{lefteqnarray}
where $M$ is the total mass and $\rho$ the local density.   The
empirical variance is:
\begin{equation}
\label{eq:spp}
\sigma_{\phi\phi}^2=\overline{(v_\phi^2)}-(\overline{v_\phi})^2~~;
\end{equation}
by definition.

The moment-of-inertia-weighted angular velocity
and squared angular velocity read:
\begin{lefteqnarray}
\label{eq:O}
&& \overline{\Omega}=\frac1{I_3}\int\int\int\overline{\Omega}(x_1,x_2,x_3,t)
w^2\rho(x_1,x_2,x_3,t)\diff x_1\diff x_2\diff x_3~~; \\
\label{eq:O2}
&& \overline{(\Omega^2)}=\frac1{I_3}\int\int\int\overline{\Omega^2}
(x_1,x_2,x_3,t)w^2\rho(x_1,x_2,x_3,t)\diff x_1\diff x_2\diff x_3~~;
\end{lefteqnarray}
where $I_3$ is the moment of inertia with respect to
the rotation axis, $x_3$.   The empirical variance is:
\begin{equation}
\label{eq:sOO}
\sigma_{\Omega\Omega}^2=\overline{(\Omega^2)}-(\overline{\Omega})^2~~;
\end{equation}
by definition.

A link between mass and moment of inertia is
provided by the following relation:
\begin{equation}
\label{eq:RG3}
R_{\rm G3}^2=\frac{I_3}M=\frac1M\int\int\int w^2\rho(x_1,x_2,x_3,t)\diff x_1
\diff x_2\diff x_3~~;
\end{equation}
where $R_{\rm G3}$ is the radius of gyration
with respect to the rotation axis, $x_3$.
Using Eqs.\,(\ref{eq:vfO}) and (\ref{eq:RG3}),
the combination of Eqs.\,(\ref{eq:vph2}),
(\ref{eq:spp}), (\ref{eq:O2}), and (\ref{eq:sOO})
yields:
\begin{lefteqnarray}
\label{eq:vOR2}
&& \overline{(v_\phi^2)}=\overline{(\Omega^2)}R_{\rm G3}^2~~; \\
\label{eq:vOR}
&& (\overline{v_\phi})^2+\sigma_{\phi\phi}^2=[(\overline{\Omega})^2+
\sigma_{\Omega\Omega}^2]R_{\rm G3}^2~~;
\end{lefteqnarray}
while, on the other hand, $\overline{v_\phi}\ne
\overline{\Omega}R_{\rm G3}$, $\sigma_{\phi\phi}
\ne\sigma_{\Omega\Omega}R_{\rm G3}$, and a different
averaging is needed on either $v_\phi$ or
$\Omega$.

To this aim, the $(MR_{\rm G3})$-weighted
circular velocity, $\overline{v_\Omega}$,
and the $(MR_{\rm G3})$-weighted angular
velocity, $\overline{\Omega_v}$, are defined as:
\begin{lefteqnarray}
\label{eq:vO}
&& \overline{v_\Omega}=\overline{\Omega}R_{\rm G3}=\frac{R_{\rm G3}}{I_3}
\int\int\int\overline{v_\phi}(x_1,x_2,x_3,t)w
\rho(x_1,x_2,x_3,t)\diff x_1\diff x_2\diff x_3;\quad \\
\label{eq:Ov}
&& \overline{\Omega_v}=\frac{\overline{v_\phi}}{R_{\rm G3}}=\frac{R_{\rm G3}}
{I_3}\int\int\int\overline{\Omega}(x_1,x_2,x_3,t)
w\rho(x_1,x_2,x_3,t)\diff x_1\diff x_2\diff x_3;\quad
\end{lefteqnarray}
and the angular momentum with respect to the
rotation axis, $x_3$, is defined by the
integral on the right-hand side of
Eq.\,(\ref{eq:O}) or (\ref{eq:vO}), which yields:
\begin{equation}
\label{eq:J3}
J_3=I_3\overline{\Omega}=M\overline{v_\Omega}R_{\rm G3}~~;
\end{equation}
in terms of the moment-of-inertia-weighted angular velocity
and the $(MR_{\rm G3})$-weighted circular velocity,
respectively.
%
%and Eqs.\,(\ref{eq:vOR2}) and (\ref{eq:vOR})
%may be cast under the equivalent form:
%\begin{lefteqnarray}
%\label{eq:vpO2}
%&& \overline{(v_\phi^2)}=\overline{(v_\Omega^2)}~~; \\
%\label{eq:vpO}
%&& (\overline{v_\phi})^2+\sigma_{\phi\phi}^2=[(\overline{v_\Omega})^2+
%\sigma_{GG}^2]~~; \\
%\label{eq:sGG}
%&& \sigma_{GG}=\sigma_{\Omega\Omega}R_{\rm G3}~~;
%\end{lefteqnarray}
%in terms of mean circular velocities.

The combination of Eqs.\,(\ref{eq:RG3}),
(\ref{eq:vOR2}), (\ref{eq:vO}), and (\ref{eq:Ov})
yields the following expression for the radius of
gyration:
\begin{equation}
\label{eq:RvO}
R_{\rm G3}^2=\frac{I_3}M=\frac{\overline{(v_\phi^2)}}{\overline{(\Omega^2)}}=
\frac{(\overline{v_\Omega})^2}{(\overline{\Omega})^2}=\frac{(\overline
{v_\phi})^2}{(\overline{\Omega_v})^2}~~;
\end{equation}
where the differences, $\overline{(v_\phi^2)}-
(\overline{v_\Omega})^2$, $\overline{(\Omega^2)}-
(\overline{\Omega_v})^2$, cannot be related
to empirical variances as the corresponding terms
are weighted in different ways.

Accordingly, the rotation kinetic energy
related to tangential equatorial velocity
components may be expressed as:
\begin{leftsubeqnarray}
\slabel{eq:Ekppa}
&& (E_{\rm kin})_{\phi\phi}=\frac12M\left[(\overline{v_\phi})^2+\sigma_
{\phi\phi}^2\right]=E_{\rm rot}+\frac12M\sigma_{\phi\phi}^2~~; \\
\slabel{eq:Ekppb}
&& E_{\rm rot}=\frac12M(\overline{v_\phi})^2~~;
\label{seq:Ekpp}
\end{leftsubeqnarray}
and the total kinetic energy reads:
\begin{leftsubeqnarray}
\slabel{eq:Ekina}
&& E_{\rm kin}=E_{\rm rot}+E_{\rm res}~~; \\
\slabel{eq:Ekinb}
&& E_{\rm res}=\frac12M\left[\overline{(v_w^2)}+\overline{(v_3^2)}+
\sigma_{\phi\phi}^2\right]~~; \\
\slabel{eq:Ekinc}
&& \overline{(v_w^2)}=(\overline{v_w})^2+\sigma_{ww}^2=\sigma_{ww}^2~~; \\
\slabel{eq:Ekind}
&& \overline{(v_3^2)}=(\overline{v_3})^2+\sigma_{33}^2=\sigma_{33}^2~~;
\label{seq:Ekin}
\end{leftsubeqnarray}
where $v_w$ and $v_3$ are the radial equatorial
and polar velocity components and $\overline{v_w}=
\overline{v_3}=0$ provided the centre of mass 
coincides with the origin of the coordinates
(for further details refer to e.g., C07).
Accordingly, Eq.\,(\ref{eq:Ekinb}) reduces to:
\begin{equation}
\label{eq:Err}
E_{\rm res}=\frac12M\sigma^2=\frac12M\left(\sigma_{\phi\phi}^2+\sigma_{ww}^2+
\sigma_{33}^2\right)~~;
\end{equation}
which is equivalent to:
\begin{equation}
\label{eq:Erc}
E_{\rm res}=\frac12M\sigma^2=\frac12M\left(\sigma_{11}^2+\sigma_{22}^2+
\sigma_{33}^2\right)~~;
\end{equation}
in Cartesian coordinates.

\section{Systematic and random rotation excess}
\label{a:rost}

Let two matter distributions be characterized
by equal density profiles and shapes, but
isotropic and anisotropic stress tensor,
respectively, and different amount of angular
momentum.   The generalized virial equations,
Eqs.\,(\ref{seq:virg}), read:
\begin{leftsubeqnarray}
\slabel{eq:visa}
&& 2[(E_{\rm rot})_{pp}]_{\rm ani}+2[(\widetilde{E}_{\rm res})_{pp}]_
{\rm ani}+(E_{\rm pot})_{pp}=0~~;\qquad p=1,2,3~~; \\
\slabel{eq:visb}
&& 2[(E_{\rm rot})_{pp}]_{\rm iso}+2[(\widetilde{E}_{\rm res})_{pp}]_
{\rm iso}+(E_{\rm pot})_{pp}=0~~;\qquad p=1,2,3~~; \\
\slabel{eq:visc}
&& (E_{\rm rot})_{11}+(E_{\rm rot})_{22}=E_{\rm rot}~~;\qquad
(E_{\rm rot})_{33}=0~~;
\label{seq:vis}
\end{leftsubeqnarray}
where $(\widetilde{E}_{\rm res})_{pp}=
\zeta_{pp}E_{\rm res}$ is the residual
kinetic-energy tensor of the related
equilibrium figure.

In both cases,
the potential-energy tensor remains
unchanged, which implies the following
relation:
\begin{equation}
\label{eq:sisa}
[(\widetilde{E}_{\rm res})_{33}]_{\rm ani}=[(\widetilde{E}_{\rm res})_{pp}]_
{\rm iso}~~;\qquad p=1,2,3~~;
\end{equation}
and the substitution of Eq.\,(\ref{eq:sisa})
into (\ref{seq:vis}) yields:
\begin{leftsubeqnarray}
\slabel{eq:eqvsa}
&& [(E_{\rm rot})_{qq}]_{\rm iso}+\Delta(E_{\rm rot})_{qq}+
[(\widetilde{E}_{\rm res})_{qq}]_{\rm iso}+\Delta(\widetilde{E}_{\rm res})_
{qq} \nonumber \\
&& \qquad=[(E_{\rm rot})_{qq}]_{\rm iso}+[(\widetilde{E}_{\rm res})_{qq}]_
{\rm iso}~~;\qquad q=1,2~~; \\
\slabel{eq:eqvsb}
&& \Delta(E_{\rm rot})_{qq}=[(E_{\rm rot})_{qq}]_{\rm ani}-
[(E_{\rm rot})_{qq}]_{\rm iso}~~; \\
\slabel{eq:eqvsc}
&& \Delta(\widetilde{E}_{\rm res})_{qq}=[(\widetilde{E}_{\rm res})_{qq}]_
{\rm ani}-[(\widetilde{E}_{\rm res})_{qq}]_{\rm iso}~~;
\label{seq:eqvs}
\end{leftsubeqnarray}
which is equivalent to:
\begin{equation}
\label{eq:eDvs}
\Delta(E_{\rm rot})_{qq}+\Delta(\widetilde{E}_{\rm res})_{qq}=0~~;
\end{equation}
where the first and the second term of the
sum may be conceived as a (positive or
negative) systematic and random rotation
excess, respectively.   Random rotation
has to be intended as related to a selected
axis, with a null mean value.

With regard to nonequilibrium figures,
the validity of the relation:
\begin{equation}
\label{eq:fDvs}
\Delta(E_{\rm rot})_{qq}+\Delta(E_{\rm res})_{qq}=0~~;
\end{equation}
means that a fixed amount of rotation
energy has been converted into residual
energy (or vice versa).   The combination
of Eqs.\,(\ref{eq:eDvs}) and (\ref{eq:fDvs})
yields:
\begin{equation}
\label{eq:efDv}
\Delta(E_{\rm res})_{qq}=\Delta(\widetilde{E}_{\rm res})_{qq}~~;
\end{equation}
and Eq.\,(\ref{eq:fDvs}) may be cast
under the explicit form:
\begin{equation}
\label{eq:efvs}
\Delta[(\overline{v_\phi})_{qq}]^2+\Delta(\sigma_{qq}^2)=0~~;
\end{equation}
according to Eqs.\,(\ref{seq:Ekpp}) and
(\ref{seq:Ekin}).   The above results may
be reduced to a single statement.
\begin{trivlist}
\item[\hspace\labelsep{\bf Theorem}] \sl
Given a matter distribution with assigned
density profile, shape, virial index,
$\zeta=\widetilde{E}_{\rm res}/E_{\rm res}$,
and isotropic stress tensor, an infinity of
adjoint configurations with anisotropic stress
tensor exist, for which the sum of systematic
and random rotation excess is null.
\end{trivlist}

According to earlier attempts (Lynden-Bell
1960, 1962; Meza 2002) the distribution
function is independent of the sign of the
tangential velocity components, and the
whole set of possible configurations is
characterized by an equal amount of both
kinetic and potential energy.   In other
words, clockwise and counterclockwise
circular motions are indistinguishable
to this respect, passing from systematic
(maximal squared mean tangential equatorial
velocity component) to
random (minimal squared mean tangential
equatorial velocity component) rotation.   An
anisotropic stress tensor could be due
to the prevalence of either systematic
rotation $(\sigma_{11}=\sigma_{22}<
\sigma_{33})$ or random rotation
$(\sigma_{11}=\sigma_{22}>\sigma_{33})$
with respect to the adjoint configuration
with isotropic stress tensor.

In the limiting case of flat configurations,
$\epsilon_{31}\to0$, $(E_{\rm pot})_{33}\to
0$ (e.g., Caimmi 2009a), Eq.\,(\ref{eq:virgb})
implies $(E_{\rm res})_{33}=\zeta_{33}
E_{\rm res}\to0$ and  $(E_{\rm res})_{qq}\to
0$ for isotropic stress tensor.   In other
words, flat configurations with isotropic
stress tensor must necessarily be self-gravitating.
Rotation curves depend on density profiles,
ranging from linear curves related to homogeneous
systems to Keplerian curves related to Roche
systems.   Accordingly, $\sigma_{\phi\phi}=0$
in the former alternative, growing up to a
maximum, $\sigma_{\phi\phi}>0$, in the latter.
But $\sigma_{\phi\phi}>0$ would imply
$(E_{\rm res})_{qq}>0$, which is in contradiction
with either sufficiently flattened configurations or isotropic
stress tensors.   Then a threshold in meridional
axis ratio exists, $\epsilon_{31}=\epsilon_{31}^
\ast$, below which configurations with isotropic
stress tensor cannot exist for an assigned
density profile.

\end{document}